\documentclass[11pt]{article}

\usepackage{fullpage,calc,amsfonts,amssymb,amsmath,bm,url,color,theorem,graphicx}
\usepackage{epsfig}
\usepackage{graphicx,color}
\usepackage{stfloats}
\usepackage{graphicx}
\usepackage{psfrag}
\usepackage{subfigure}
\usepackage{url}
\usepackage{stfloats}
\usepackage{amsfonts,amssymb,amsmath,bm,paralist,theorem,ifthen,color}
\usepackage{array}
\usepackage{caption}
\usepackage{calc}
\usepackage{subfig}
\usepackage{longtable}
\usepackage{stfloats}
\usepackage{graphics,color,epsfig,subfigure,enumerate}
\usepackage{multirow}
\usepackage{algorithmic,algorithm}
\usepackage{setspace}
\usepackage[style=nature]{biblatex}
\usepackage{comment}
\usepackage{authblk}
\usepackage{placeins}
\usepackage{mathtools}
\newtheorem{Lemma}{Lemma}

\theorembodyfont{\rmfamily}

\graphicspath{{Figures/}}
\title{HYPERION: Hyperspectral Penetrating-type Ellipsoidal Reconstruction for Terahertz Blind Source Separation}

\author[1,2,6]{Chia-Hsiang Lin}
\author[3,6]{Yi-Chun Hung}
\author[4,5]{Feng-Yu Wang}
\author[4,$\dagger$]{Shang-Hua Yang}
\affil[1]{Department of Electrical Engineering, National Cheng Kung University}
\affil[2]{Miin Wu School of Computing, National Cheng Kung University}
\affil[3]{Department of Electrical and Computer Engineering, University of California, Los Angeles}
\affil[4]{Department of Electrical Engineering, National Tsing Hua University}
\affil[5]{Department of Life Science, National Tsing Hua University}
\affil[6]{Contributed equally}

\begin{document}
\maketitle
\section{\Large{Abstract}}
Terahertz (THz) technology has been a great candidate for applications, including pharmaceutic analysis, chemical identification, and remote sensing and imaging due to its non-invasive and non-destructive properties. Among those applications, penetrating-type hyperspectral THz signals, which provide crucial material information, normally involve a noisy, complex mixture system. Additionally, the measured THz signals could be ill-conditioned due to the overlap of the material absorption peak in the measured bands. To address those issues, we consider transmitted (penetrating-type) signal mixtures and aim to develop a \textit{blind} hyperspectral unmixing (HU) method without requiring any information from a prebuilt database, such as predefined material composition or material information. The proposed HYperspectral Penetrating-type Ellipsoidal ReconstructION (HYPERION) algorithm is unsupervised, not relying on collecting extensive data or sophisticated model training. Instead, it is developed based on elegant ellipsoidal geometry under a very mild requirement on data purity, whose excellent efficacy is experimentally demonstrated.

\section{\Large{Introduction}}
The recent advances in remote sensing technologies have revolutionized the ways people interact with the world in communication, manufacturing, molecular science, life science, healthcare, geography, and astronomy \cite{koenig2013wireless, curran1987review, akiyama2019first}. Among all remote sensing methods that perceive information non-invasively, electromagnetic (EM) wave has become one of the most critical tools due to their promising nature of the high-speed operation, long traveling distance, and unique light-matter interactions in specific spectral bands \cite{gorelick2017google}. With the use of the EM wave, unveiling multi-functional behaviors inside objects with deep-subwavelength spatial precision and tera-frame-per-second imaging speed has also been well-demonstrated nowadays \cite{faccio2018trillion}. Terahertz (THz) spectrum, located between microwave and infrared EM spectrum, has recently aroused extensive attention for identifying a great variety of materials, including molecules, proteins, explosives, chemical mixtures, and charged particles in a remote distance \cite{globus2003thz, markelz2007protein, hung2020penetrating, george2008ultrafast}. The chemical compositions and the molecular structures of materials can directly map their unique molecular dynamics in the THz frequency regime. This enables the THz spectroscopy tools to unveil fruitful material information inside optically opaque objects without any labels. Due to the non-ionizing, non-destructive nature of THz wave, THz spectroscopy systems are of great interest for medical, pharmaceutical, non-destructive evaluation, and industrial inspection \cite{kawase2003non, federici2005thz}. Different systems based on THz spectroscopic modalities have further extended the application fields. THz near-field systems such as THz scanning near-field microscopy \cite{huber2008terahertz} and THz scanning tunneling microscopy \cite{cocker2013ultrafast} significantly shrink spatial resolution down to atomic level, which becomes powerful tools to perform near-field studies of advanced materials and resolving structural information of biological samples \cite{cocker2013ultrafast}. Material dynamics down to tens of femtosecond temporal resolution can be further determined through time-resolved THz spectroscopy \cite{schmuttenmaer2004exploring}. In addition, hyperspectral THz imaging systems map out not just water content inside biological samples but chemical distribution inside a sealed package \cite{kawase2003non, federici2005thz}. 

Although THz spectroscopy has shown its great promises in the past decades, many materials, chemical compounds, and biomolecules have not yet been fully characterized in the THz frequency range, unlike in visible light and infrared spectrum. Furthermore, different THz spectroscopic modalities, measuring conditions, or sample preparation methods would introduce considerable discrepancies between samples with the same material \cite{cherkasova2020thz, upadhya2003terahertz}. The typical way to perform THz spectroscopy for material identification is to establish THz material characteristics of pure substances locally, followed by material information extraction \cite{dorney2001material}. This way, a ppb-level sensing capability has been demonstrated by evaluating the spectral shifts, and amplitude changes of the material absorption spectrum in the THz regime \cite{zhong2006identification}. However, in many real-world scenarios, such as identifying chemical portions of malignant tumors or \textit{in situ} quantifying glucose in a body fluid sample, it is still challenging to purify and extract pure substances for further THz material dataset establishment. This physical constraint significantly limits the practical use of THz spectroscopy systems in a wide range of applications. We want to ask here: \textit{Can we separate material signatures of mixture chemicals without measuring or even knowing their pure substances?} This is a blind spectral unmixing problem waiting to be answered, especially in THz spectroscopy, THz hyperspectral imaging, and THz biophotonics fields. Recently, research groups have implemented THz spectral unmixing methods for material blind separation – hard modeling factor analysis (HMFA) \cite{li2015component}, nonnegative matrix factorization (NMF) \cite{ma2017thz}, and independent component analysis (ICA) \cite{balci2016independent}. However, all the conventional THz spectral unmixing methods still rely on properties of material information, the size of a measured dataset, and the signal-to-noise ratio of the dataset, which severely limits their scope of practical use. 
HMFA uses the correlation of mixture signals to unmix sources, which requires curve fitting for the massive amount of material information, such as spectral peak amplitudes, peak locations, and spectral profiles. 
The blind unmixing performance of the nonconvex NMF is sensitive to initial parameters and the signal-to-noise ratio, leading to difficulties for practical uses. 
One of the requirements in ICA is the statistical independence of the sources, causing inaccurate amplitudes of THz unmixed spectrum compared with the ground truth. 
Here, we proposed an unsupervised blind separation algorithm, HYperspectral Penetrating-type Ellipsoidal ReconstructION (HYPERION), which needs neither collecting big data nor sophisticated model training. In this article, we demonstrated a comprehensive study on HYPERION for non-invasive, non-destructive material separation and identification. To the best of our knowledge, HYPERION has shown superior performance among the state-of-the-art THz spectral unmixing methods regarding separation accuracy (i.e., spectral angle error, mean square error) and noise immunity testing with mixture materials.   Due to the superior blind separation capability, we have further demonstrated highly accurate material blind separation mapping through HYPERION THz hyperspectral imaging. It should be noted that there is no prior knowledge of pure substance information, no supervised model training, no need to establish THz material dataset for pure substances, and no restricted requirement on the data purity $\gamma$. Our proposed blind separation modality, HYPERION, could potentially make paradigm shifts of THz technology for chemical sensing, industrial inspection, material identification, biomedical sensing, and imaging in the near future.

\section{\Large{Results}}
\subsection*{\large{Assumption evaluation}}

Our method, HYPERION, is designed based on L\"owner-John ellipsoid (LJE) and the linear mixing model ((Supplementary Note 14)), where the THz data follows a convex geometry structure (see ``Methods" for the details). By utilizing the information of the LJE defined as the maximum-volume ellipsoid inscribed in the convex geometry, HYPERION has the efficacy to blindly unmix sources from the data with low data purity \cite{lin2018maximum}. To validate the linear mixing model on the transmitted THz signals, we have measured and characterized mixture tablets composed of D-lactose monohydrate, D-sucrose, and L-tyrosine through a broadband THz time-domain spectroscopy (THz-TDS) system. The three chemicals are selected due to their abundant information within THz spectral bands, including spectral absorption lines and varying absorption spectral trends. The obtained ternary dataset contains the three pure-substance THz spectra (D-lactose monohydrate, D-sucrose, and L-tyrosine) and six mixture spectra composed by each pair of the chemicals with a ratio of 3:7, which is designed to test the data purity criterion. The details of measurement protocol and environment condition are elaborated in Supplementary Note 1.

\begin{figure}
  \centering
  \includegraphics[width=0.75\linewidth]{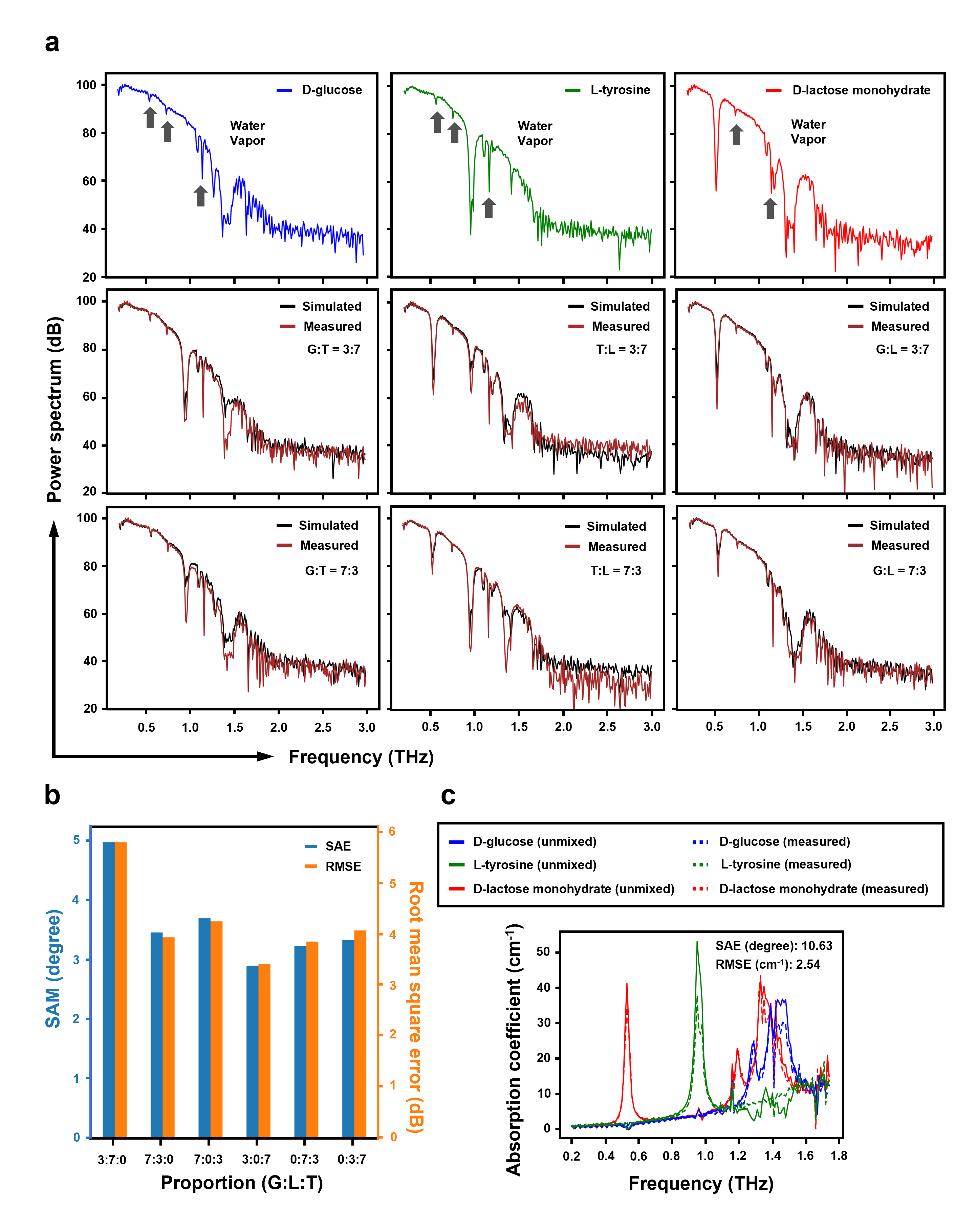}
  \captionsetup{font={small,stretch=1.25}}
  \caption{\textbf{Unmixed results of the ternary dataset by HYPERION.} \textbf{(a)} The three subplots at the top panel are the measured spectra of pure substances (D-glucose, L-tyrosine, and D-lactose monohydrate). Gray arrows indicate the absorption lines of water vapor. The bottom two rows of subplots are the mixture spectra with different mixing ratios; D-glucose, D-lactose monohydrate, and L-tyrosine are denoted as G, L, and T, respectively. The simulated lines (black) are computed based on the linear mixing model and the measured spectra of the pure substances. \textbf{(b)} The spectral angle mapper (SAM) and root mean square error (RMSE) between the simulated and measured spectra. G:L:T represents the proportion of D-glucose, D-lactose monohydrate, and L-tyrosine. \textbf{(c)} The comparison between the unmixed material absorption spectrum and the measured absorption spectrum upon the ternary dataset with pure substances. The unmixed and measured material absorption spectrum are shown as solid lines and dotted lines, respectively.}
  \label{fig:1}
\end{figure}

We compared the THz spectra simulated by the convex combinations of the three chemicals with their corresponding THz measured spectrum with the ternary dataset. All samples are prepared based on the standardized protocol (see ``Methods" for the details). As shown in Fig. \ref{fig:1}(a), all the simulated spectra are highly aligned with the measured spectra, including the spectral features and the locations of spectral lines. We then project high dimensional data of the ternary dataset into a low dimensional space through principal component analysis (PCA). The PCA evaluation result provides solid evidence that the necessary criteria of holding the linear mixing model are well-preserved (Supplementary Note 2). To further explore the linear mixing model in terms of spectrum similarity and accuracy, we use spectral angle mapper (SAM) and root-mean-square error (RMSE) \cite{wei2015hyperspectral, loncan2015hyperspectral} to evaluate the diminutive differences between simulated spectrum and measured spectrum (Supplementary Note 3). The SAM indicator is based on the projection angle between two spectra in comparison, which explicitly reveals the spectral shape similarity without mixing the influence of spectrum offset. The RMSE indicator is the square root of the quadratic mean of differences between the two spectra. It includes both shape similarity and the effects of the band-wise differences among the measured spectral range. The SAM and RMSE of this ternary dataset are less than 5 degrees and 6 cm$^{-1}$ as shown in Fig. \ref{fig:1}(b), which demonstrates both the excellent spectral shape similarity as well as the slight deviation between the measured and simulated mixture spectra. Based on the validation of qualitative and quantitative analysis, the assumption of the linear mixing model is experimentally verified. Note that the linear mixing model has been extensively used in remote sensing for reflected (reflecting-type) signals but never for transmitted (penetrating-type) signals as far as we know; thus, the experiment is needed. We then transform the ternary dataset from the THz transmission spectrum to the material absorption spectrum of each measured sample (Supplementary Note 4), which directly presents broadband material information, including resonant frequencies, spectral features, and light-matter interaction levels in the THz range. Additionally, we extract and present material absorption spectrum between 0.2 THz to 1.75 THz since the ASOPS THz-TDS system dynamic range in power spectra of the selected materials approximately decreases to about 5 dB at 1.75 THz that start to mismatch with the linear mixing model. As shown in Fig. \ref{fig:1}(c), all unmixed pure substance absorption spectra are highly aligned with the measured ones, including the spectral peak locations and the material absorption profiles. The unmixed D-glucose and L-tyrosine values are slightly deviated from the measured values above 1.5 THz since the signal-to-noise ratio (SNR) of the THz spectroscopy system gradually decreases from 60 dB at 0.3 THz to 5 dB at 1.75 THz.

Nevertheless, at low-SNR spectral regions (1.2 - 1.6 THz), HYPERION still provides great unmixing capability - less than 3.51 cm$^{-1}$ deviated absorption coefficient values from the measured ones. For the spectral regimes with a noticeable unmixing difference, for example, the measured values of L-tyrosine absorption peak are 15.35 cm$^{-1}$ and 4.78 cm$^{-1}$ differences from the unmixed values at 0.949 THz and 1.295 THz, respectively. Due to the overlap of water vapor and material absorption peaks, the linear mixing model is not well-preserved. Since HYPERION takes all the spectral information into account by the LJE instead of specific spectral components, even in nonlinear matter-matter or light-matter interaction spectral regions, the unmixed material absorption spectrum based on HYPERION still demonstrates great spectral accuracy of each source spectral signature but just a relatively higher mismatch in absorption coefficient values.

Despite the nonlinear interaction and the low SNR levels in certain bands, HYPERION offers an accurate unmixing performance – 10.63 degrees with SAM and a 2.54 cm$^{-1}$ with RMSE of overall unmixed material absorption spectrum has been achieved. This SAM value presents high spectral shape similarity of unmixed absorption coefficients, preserving rich material spectral profiles and details among a wide frequency range. In another perspective, the extraordinarily low RMSE shows both high unmixing accuracy and spectral similarity of the unmixed absorption coefficient, providing various types of material spectral features for further chemical identification and non-destructive evaluation applications.

\subsection*{\large{Quinary Dataset with Pure Substances}}

HYPERION has several advantages, including support of complex mixture systems, milder data purity requirement, excellent spectral accuracy on unmixed sources, and remarkable noise immunity, due to the simultaneous use of all spectral information by LJE. To further investigate the performance of HYPERION in broad application scope, we design three experimental modules to verify HYPERION characteristics in terms of unmixing capability of complex mixture system, low data purity dataset, and noisy environment condition based on the quinary dataset (see ``Methods" for the details). The quinary dataset contains the five pure chemicals (D-glucose, D-lactose monohydrate, L-tyrosine, L-histidine, and D-sucrose) with rich material information in THz spectral range and ten mixture tablets. The five pure chemicals are specifically selected to validate HYPERION performance under different unmixed spectral scenarios - identifying single/multiple spectral peaks and overlaid spectral profiles simultaneously. Within the spectral range of 0.2 - 1.7 THz, L-tyrosine and L-histidine have a single absorption peak at 0.95 and 0.78 THz, respectively. D-lactose monohydrate contains two absorption peaks located at 0.52 THz and 1.3 THz. D-sucrose and D-glucose present the high spectral similarity of their THz absorption spectra, which is used to evaluate ill-conditioned levels of HYPERION.  Furthermore, D-lactose is also selected to investigate the unmixing performance of HYPERION on the pharmaceutical application since D-lactose is the common ingredient as an excipient in tablets \cite{nowak2015selected,gamble2010investigation}. The ten mixture tablets are composed of each pair of the five chemicals with a 5:5 ratio, which satisfies the milder requirement of the data purity for a quinary case \cite{lin2014identifiability}. Identical experimental setup, tablet preparation, data acquisition, and data preprocessing protocol are well-followed as the previous ternary dataset establishment.

\begin{figure}
  \centering
  \includegraphics[width=0.8\linewidth]{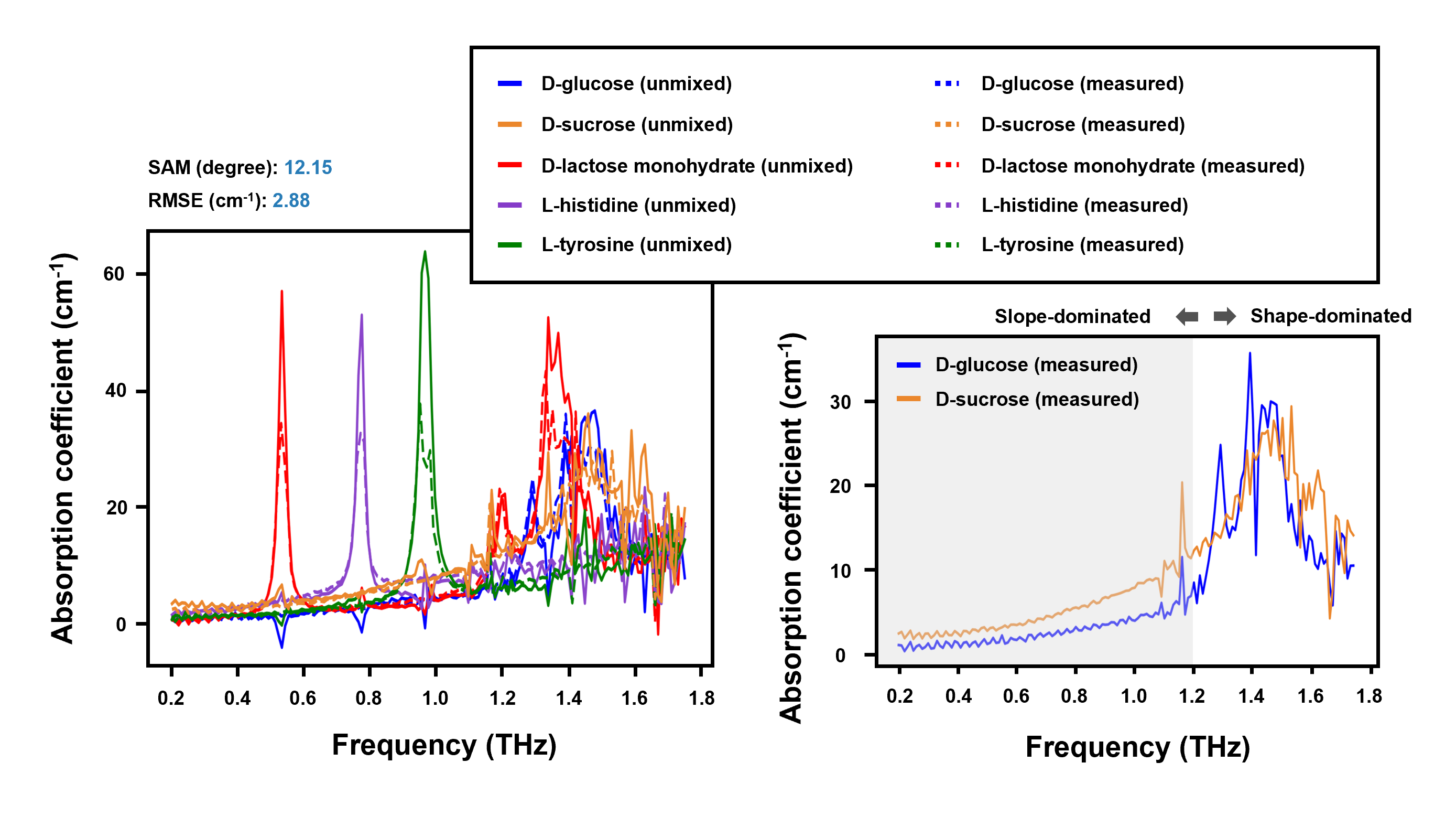}
  \captionsetup{font={small,stretch=1.25}}
  \caption{\textbf{Unmixed material absorption spectra of the quinary dataset with pure substances.} The comparison between unmixed material absorption spectrum (solid line) and measured material absorption spectrum (dotted line) upon the quinary dataset with pure substances. The subset of the figure shows that the D-glucose and D-sucrose absorption spectra are highly correlated. The material absorption spectrum difference between D-glucose and D-sucrose within 0.2 - 1.2 THz is dominated by different slopes (gray region) while it is dominated by shape within 1.2 - 1.75 THz.}
  \label{fig:2}
\end{figure}

Comparing the unmixed and measured absorption spectra shows that HYPERION has great efficacy on the different spectral scenarios. As shown in Fig. \ref{fig:2}, the five unmixed absorption spectra (solid lines) based on the quinary dataset demonstrate a high correlation with the measured absorption spectra (dotted lines) within a broad frequency range. In terms of spectral peak locations, unmixed absorption spectra with varieties of material properties are precisely resolved. The spectral peak deviation of all unmixed absorption spectra is less than 10 GHz, close to the spectral resolution limit of the asynchronized optical sampling (ASOPS) THz-TDS system (see ``Methods" for the details) at protocol settings. Additionally, the increasing trends and the characteristics of the absorption spectra are well unmixed since HYPERION utilizes the convex geometry of the whole measured frequency bands for the blind separation instead of specific bands information, such as peak regimes. Those mismatched values between unmixed and measured absorption spectra mainly come from nonlinear light-matter interaction \cite{heshmat2017terahertz}, and limited SNR levels at higher frequencies.

To address the ill-conditioned case, HYPERION transforms the data matrix into the preconditioned space, where the similar material absorption spectra can be easily separated (see ``Methods" for the details). As shown in Fig. \ref{fig:2}, D-glucose and D-sucrose absorption spectra are highly correlated except for the slightly different slope within the 0.2 THz – 1.2 THz frequency range. Under this condition, HYPERION again shows the unmixing capability of a SAM of 9.57 degrees and an RMSE of 2.30 cm$^{-1}$, demonstrating its resilience to ill-conditioned scenarios with the complex mixture system. Considering different spectral characteristics among the quinary dataset, HYPERION still reaches 12.15 degrees and 2.88 cm$^{-1}$ in SAM and RMSE, respectively. Compared with less complicated material systems (e.g., ternary system), HYPERION maintains a similar accuracy level of unmixed material spectrum based on complex mixture systems (e.g., quinary system). Additionally, HYPERION is featured for its low computation time due to the convexity nature of the LJE (see ``Methods" for the details). The computation time of HYPERION on this dataset is less than 3 seconds under a general personal laptop (Intel Core i5 with 8 GB memory).

\subsection*{\large{Quinary Dataset without Pure Substances}}

In most on-site blind source separation scenarios, such as pharmaceutical inspection, biomarker analysis, remote sensing, and chemical identification, the existence and identification of pure substances remain unknown, which generally involve highly mixed chemicals and imply a low data purity. To this end, it is crucial to extract source signals from mixtures with low data purity \cite{lin2018maximum}.

\begin{figure}
  \centering
  \includegraphics[width=0.8\linewidth]{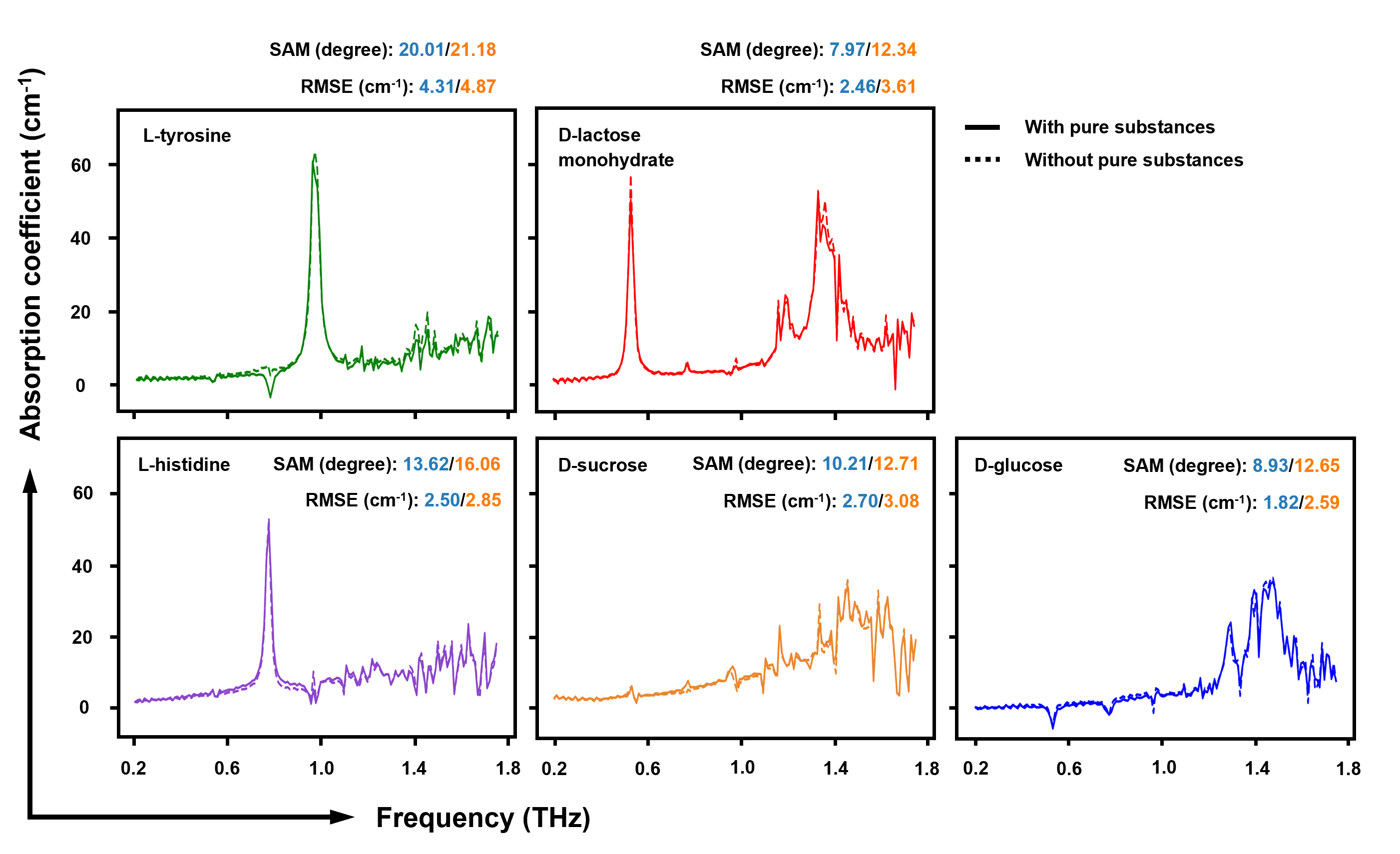}
  \captionsetup{font={small,stretch=1.25}}
  \caption{\textbf{Comparison of unmixed material absorption spectra with and without pure substances of the quinary dataset.} Quinary dataset (D-glucose, D-lactose monohydrate, D-sucrose, L-tyrosine, L-histidine) with and without pure substances unmixed by HYPERION are indicated by solid and dotted lines, respectively. The values of spectral angle mapper (SAM) and root-mean square error (RMSE) of dataset with and without pure substances are also shown (with/without) to evaluate the spectral shape similarity.}  
  \label{fig:3}
\end{figure}

Here, we have evaluated the HYPERION based on the same quinary dataset but excluded all pure substances. There is no prerequisite information from all mixture tablets in this testbed, including spectral information, the existence of pure substances, and spectra composition. To address this type of dataset with scarce information, HYPERION fits a simplex to the data convex geometry in the preconditioned space. When the optimal simplex is constructed, the corners of the optimal simplex correspond to the unmixed sources (viewed as vectors) in the preconditioned space (see ``Methods" for the details). As shown in Fig. \ref{fig:3}, the difference of unmixed absorption spectra among cases with and without pure substances is almost negligible. This is because LJE has been demonstrated to provide comparable performance with the pure substance case \cite{lin2020nonnegative}, as long as the requirement of data purity is satisfied (Supplementary Note 5). The slight difference in performance comes from the measurement noise of the dataset and the number of measured datasets – 15 versus 10 measured tablets.  Without the need for pure substance information and its measurement on-site, HYPERION has the potential to expand the THz spectroscopy application scopes, opening up the door to industrial applications in non-invasive sensing, chemical identification, and \textit{in vivo} biomarker extraction.

\subsection*{\large{Noise Immunity Evaluation}}

\begin{figure}
  \centering
  \includegraphics[width=0.8\linewidth]{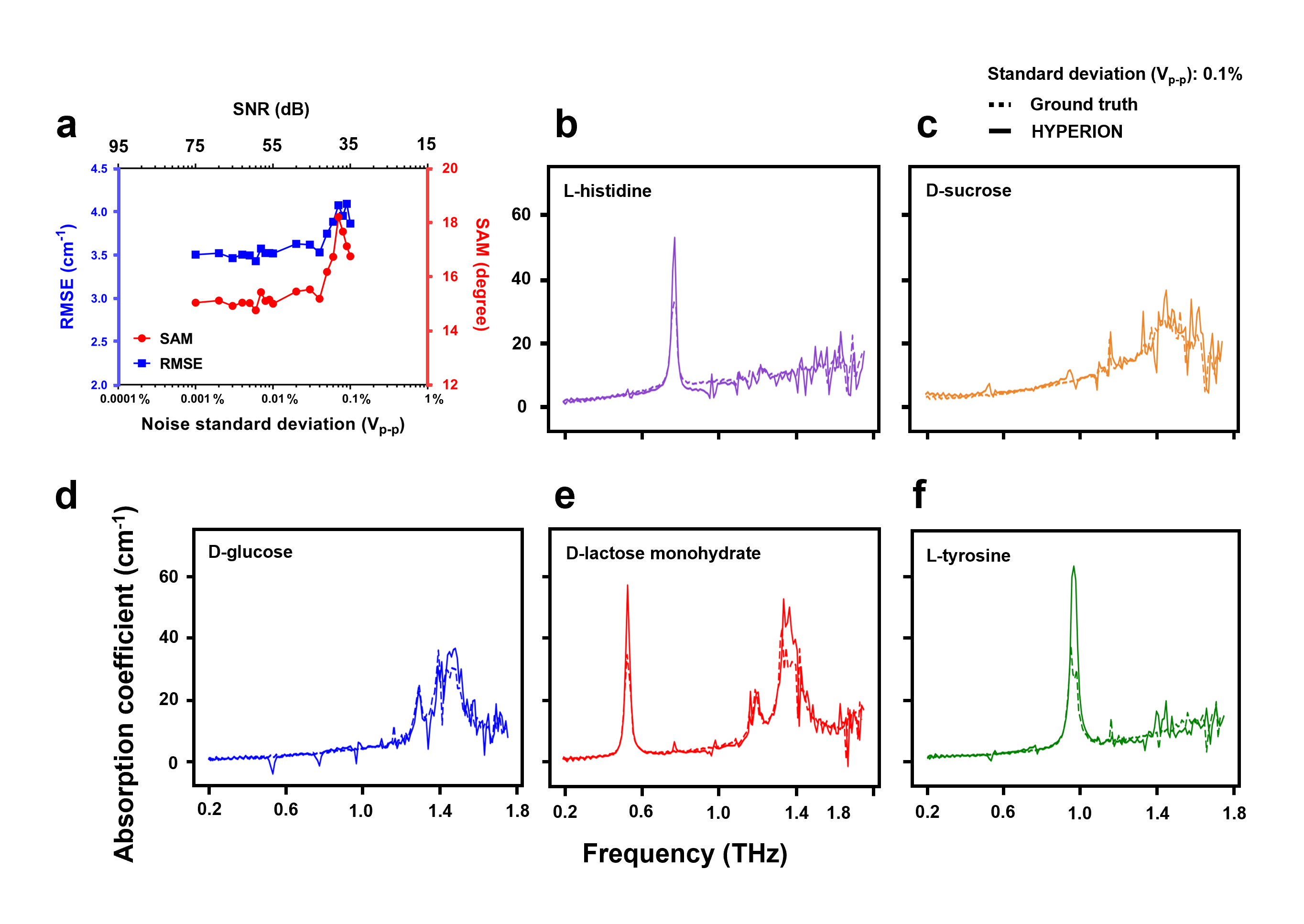}
  \captionsetup{font={small,stretch=1.25}}
  \caption{\textbf{The performance of HYPERION under different noise standard deviation (SD). (a)} The root mean square error (RMSE) and spectral angle mapper (SAM) of HYPERION under different noise SD. \textbf{(b-f)} The unmixed absorption spectra and ground truth material absorption spectra of L-histidine, D-sucrose, D-glucose, D-lactose monohydrate, and L-tyrosine within 0.2 - 1.75 THz under SD = 0.1\%.}
  \label{fig:4}
\end{figure}

To evaluate the efficacy of HYPERION with THz spectroscopy systems in general, we further investigated the noise immunity performance of HYPERION. In this study, we apply additive white Gaussian noise (AWGN) to the measured THz time-domain electric field signals with a 0.001\% - 0.1\% standard deviation (SD) range (Supplementary Note 10). Moreover, the quinary THz dataset without pure substances is chosen to better evaluate real-world scenarios – low data purity and noisy condition. In those conditions, the data convex geometry will be distorted, resulting in the inaccuracy of unmixed material absorption spectra.
To overcome the issue, HYPERION includes a general regularizer in the objective function to accommodate the distortion of convex geometry. With the general regularizer and the regularization parameter, $\lambda$, HYPERION does not force unmixed spectra to form the theoretically required regular structure in the noiseless case. Instead, it encourages the regular structure (see ``Methods" section for the details). Based on the design of the regularizer, HYPERION is well-suitable for the penetrating-type THz data, which is normally ill-conditioned and noisy. As shown in Fig. \ref{fig:4}(a), the SAM and RMSE of HYPERION unmixing results maintain at low levels while the noise SD is less than 0.1\%. In the cases of noisy environment (e.g., top six highest SD conditions), HYPERION performs slightly inferior since the SNR of THz signals in the high-frequency regime drop dramatically at higher noise levels, which causes severe distortion of the data convex geometry and leads to a deteriorating impact of the unmixing performance. To further evaluate the noise immunity capability of HYPERION among unmixed chemicals and frequency ranges, the unmixed absorption spectra of the five pure substances are demonstrated at 0.1\% noise SD (Fig. \ref{fig:4}(b)-(f)). Under this severe noise condition, the unmixed absorption spectra through HYPERION still show great alignment with the noise-free ground truth in low-frequency bands. Within the 0.2 – 1 THz frequency range, the SAM and RMSE of the D-glucose absorption spectrum are 12.32 degrees and 2.50 cm$^{-1}$, respectively. In higher frequency bands, a noticeable difference of every material absorption spectrum starts to show. Although the global trends and features are mostly well-preserved, the high-frequency spectral fluctuations due to the severe additive noise could be an issue for high-precision chemical identification applications.

In sum, HYPERION demonstrates its noise immunity feature with distorted convex geometry datasets. By utilizing this powerful feature, HYPERION not only can fit in a strictly controlled data acquisition environment but also be well suitable for real-world blind detection schemes under a noisy environment.

\subsection*{\large{Comparison with commonly used unmixing methods}}

\begin{figure}
  \centering
  \includegraphics[width=0.8\linewidth]{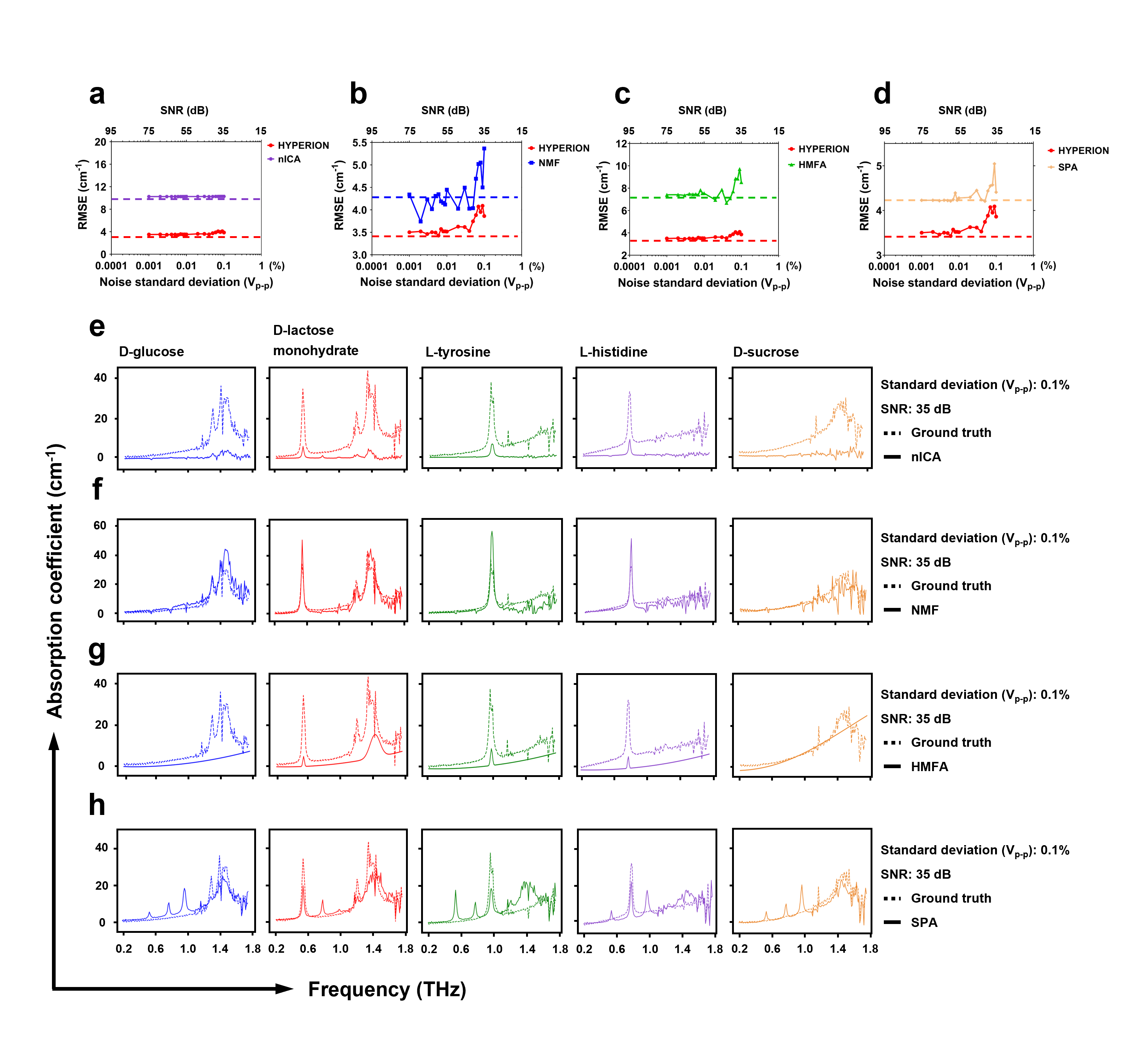}
  \captionsetup{font={small,stretch=1.25}}
  \caption{\textbf{The performance comparison among HYPERION, NMF, HMFA and SPA under different noise SD. (a-d)} The performance comparison between HYPERION with NMF, nICA, HMFA and SPA under different noise SD, respectively. The noise-free values are indicated with dotted lines. The unmixing performance is evaluated by root mean square error (RMSE) between the unmixed absorption spectra and the ground truth. \textbf{(e-h)} Five unmixed absorption spectra by NMF, nICA, HMFA and SPA are compared with the ground truth.}
  \label{fig:5}
\end{figure}

To further evaluate the unmixing performance of the geometry-based HYPERION, we introduce the commonly used unmixing methods as a comparison: statistic-based nonnegative independent component analysis (nICA), algebra-based nonnegative matrix factorization (NMF), model-based hard modeling factor analysis (HMFA), and geometry-based successive projection algorithm (SPA) \cite{berry2007algorithms, plumbley2003algorithms, ma2017thz, kriesten2008identification, balci2016independent, arora2013practical}. nICA is one of the modified versions of ICA, where it imposes the nonnegative constraint on unmixed sources. With the nonnegative constraint, nICA can unmix the spectra of nonnegative values and lead to better performance than ICA since the fraction of incident radiation absorbed by the material is always nonnegative, leading to the nonnegative values of a material absorption spectrum. NMF is the unmixing method for blind separation problems designed according to the non-convex optimization approaches (Supplementary Note 13). HMFA is the unmixing method based on peak fitting and has demonstrated its efficacy in mid-infrared and THz bands \cite{kriesten2008identification, li2015component}. SPA uses the selection and projection in the vector space to unmix the blind sources. The comparison among HYPERION, nICA, NMF, HMFA, and SPA under different noise SD is shown in Fig. \ref{fig:5}. From the unmixed material spectra and the RMSE values, HYPERION has outperformed the THz unmixing methods under a broad noise SD range from 0.001\% to 0.1\%. As expected, it comes from the three designed properties of HYPERION. First, HYPERION filters out most noise in the affine fitting step, where the data is projected into the lower dimensional convex hull by the singular value decomposition approach. Second, HYPERION uses convex geometry to find the LJE, which can further address the ill-conditioned nature of THz source signals. Third, the embedded general regularizer provides better accommodation to the noisy data by soft fitting for the distortion of convex data geometry in the objective function. As shown in Fig. \ref{fig:5}(a, e), although nICA can resolve the spectral absorption peaks, the unmixed material absorption spectra by nICA are severely compressed.
The reason is that the nICA method assumes all material absorption spectra are independent in the complex mixture system \cite{plumbley2003algorithms}. However, the summation of the chemical composition (in \%) must equal to one, making it impossible for the sources to be statistically independent, and this sum-to-one constraint leads to an assumption of nICA independent assumption.
In Fig. \ref{fig:5}(b), NMF shows decent unmixing capability because it can converge to the local optimum \cite{lin2007convergence, albright2006algorithms}. However, NMF is relatively sensitive to noisy conditions, probably due to its non-convexity nature. As a result, the unmixing performance of NMF on the THz dataset bounces dramatically at different unmixing trails and noise levels. Consequently, it is challenging to accurately resolve unmixing material spectra under a low SNR spectral regime. As shown in Fig. \ref{fig:5}(f), the unmixed absorption spectra through NMF deviate more from the ground truth signal compared to HYPERION while frequency increases. In the case of HMFA (Fig. \ref{fig:5}(c, g)), it is not capable of resolving the ill-conditioned data (i.e., D-Glucose and D-Sucrose) nor the spectral trends since the peak fitting approach inherently has difficulties in differentiating overlapped features. As shown in Fig. \ref{fig:5}(d, h), SPA has the inferior unmixing performance in the absorption peak regions since the pure-pixel assumption does not hold in those datasets \cite{ma2013signal}. The convergence time of these methods is also provided in Supplementary Table. 8. Overall, HYPERION delivers superior performance than the four THz unmixing modalities in resolving broadband spectral features, unmixing ill-condition data, reducing noise impact, and preserving spectral details in a noisy environment.

\subsection*{\large{Application}}
\begin{figure}
  \centering
  \includegraphics[width=0.75\linewidth]{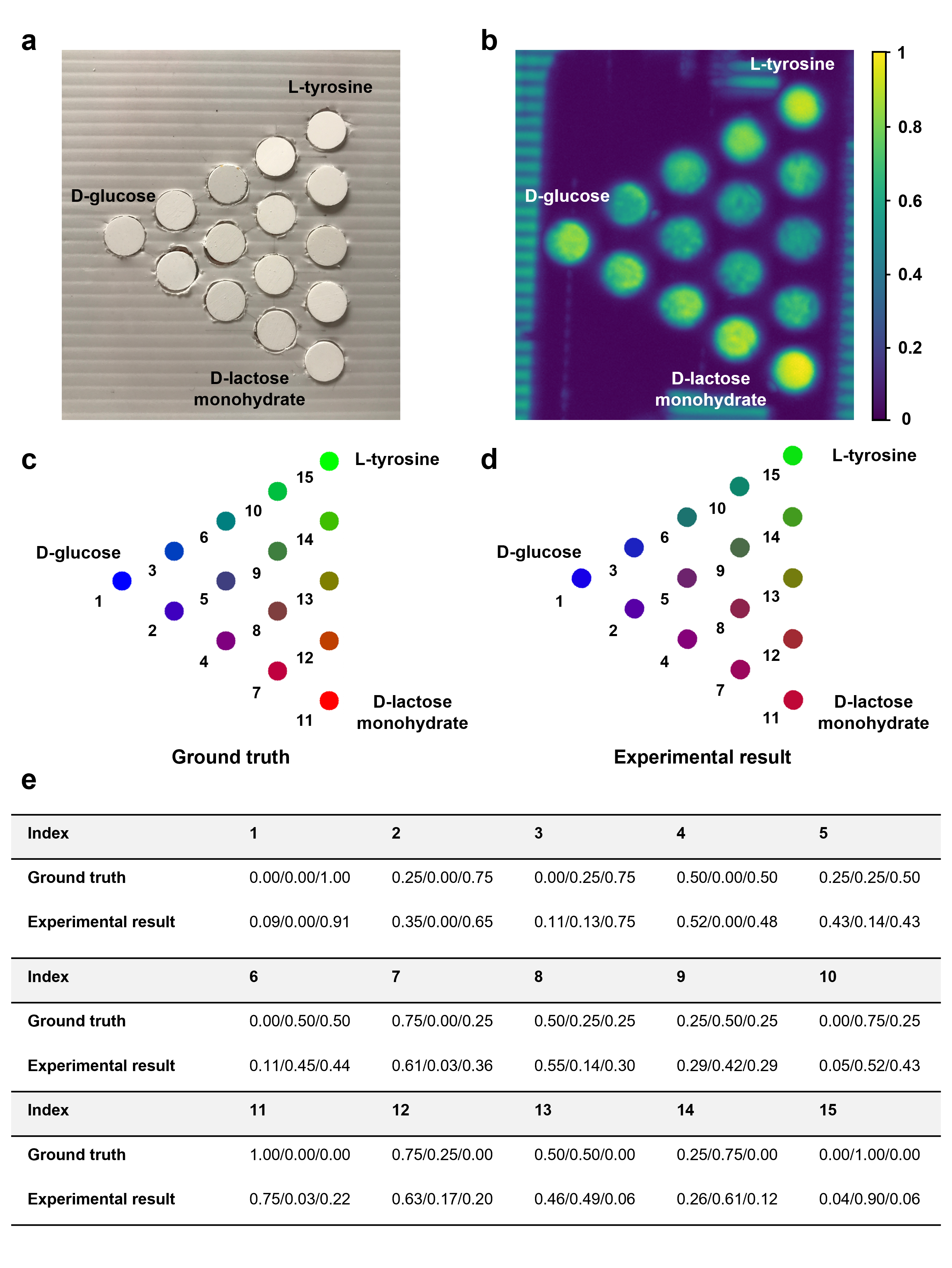}
  \captionsetup{font={small,stretch=1.25}}
  \caption{\textbf{Blind chemical mapping by THz HYPERION. (a)} The optical and \textbf{(b)} THz images of 15 tablets with different proportions composed of D-glucose, D-lactose monohydrate, and L-tyrosine, respectively. The THz image is scanned with the step size of 0.25 mm; the colormap shows the normalized field strength of the time-resolved transmitted THz signal. \textbf{(c)} The ground truth and \textbf{(d)} the experimental compositions of tablets. The color basis is blue, red, and green as shown in the ground truth $1^\text{st}$, $11^\text{th}$, and $15^\text{th}$ points, respectively. The color channels are linear combination of the selected color basis by the corresponding ratio of the chemical composition. $1^\text{st}$, $11^\text{th}$, and $15^\text{th}$ points are corresponding to D-glucose, D-lactose monohydrate and L-tyrosine, respectively. \textbf{(e)} The comparison between the estimated proportions by THz HYPERION and the ground truth (D-lactose monohydrate/L-tyrosine/D-glucose).}
  \label{fig:6}
\end{figure}

In the sections above, HYPERION has demonstrated superior features on unmixing penetrating-type THz material spectra. Those accurately unmixed spectra can benefit several applications. To demonstrate one of the potential applications, we propose the combination of the raster-scanning THz-TDS system and HYPERION (THz HYPERION) to visualize the ``secret recipes" of arbitrary objects at a remote distance. We use the unmixed spectra from the quinary dataset without pure substances and prepare a small test set of 15 chemical tablets with different compositions based on D-lactose monohydrate, D-glucose, and L-tyrosine. The 15 tablets are mounted on the paper board covered by a copper foil, which is utilized to block THz signals encoding the material information of the paper board. Every tablet is measured by the THz-TDS system with the configurations described earlier. The images in visible light and THz band images of the test set are shown in Fig. \ref{fig:6}(a) and \ref{fig:6}(b), respectively. The unmixed material spectra by the quinary dataset without pure substances are used to restore the chemical compositions of the 15 tablets in the test set by the following convex optimization problem:
\begin{equation}
\begin{aligned}\label{eq:application}
	\min_{\bm{r}_i\in\mathbb{R}^{3}}~&~\left\| \bm{x}_i - \bm{A}\bm{r}_i\right\|_1
	\\
	\textrm{s.t.}~&~
	\ \bm{r}_i \geq 0,
	\ \bm{1^T}\bm{r}_i = 1, 
\end{aligned}
\end{equation}
where $i=1, 2, \dots, 15$. $\bm{x}_i$ and $\bm{r}_i$ are the measured THz material absorption spectra of the test set and the optimization variables to estimate the corresponding ratio of compositions, respectively. 
Columns of $\bm{A}$ are the unmixed spectra from the quinary dataset without pure substances. The $L^1$ norm is chosen to effectively decrease the influence of the unmixed spectra deviation in the absorption peak regions. Since equation (\ref{eq:application}) is a convex problem, the solver CVX \cite{grant2008cvx} is adopted for the following demonstration (Supplementary Note 6). 
The qualitative comparisons between the ground truth and the estimated material composition map are shown in Fig. \ref{fig:6}(c, d). 
As expected, the majority of tablet material compositions are highly aligned with the ground truth except very few tablets (i.e. 10$^\text{th}$, 11$^\text{th}$ and 12$^\text{th}$ tablets). 
This experimental result indicates that THz HYPERION is applicable for estimating the arbitrary composition of complex mixture systems while the pure material
spectra are accurately unmixed.
In Fig. \ref{fig:6}(e), we further evaluate the unmixed material composition map qualitatively. Most unmixed results present less than 20\% inaccuracy of each composition difference compared with ground truth. The inaccuracy of estimating material composition is caused by the number of spectra in the quinary dataset, the high absorption spectrum similarity of D-lactose monohydrate and D-glucose, and the increased system noise in the high-frequency regime. It is worth mentioning that the accuracy of unmixed spectra would be greatly enhanced while the size of the measured dataset is increased. To this extent, the quality of unmixed material signatures can be significantly elevated by extending the measurement area and the pixel number of the multi-material objects under test. Furthermore, different regularizers can be further designed to match with HYPERION for specific application purposes. Overall, our demonstration shows the potential for some blind separation mapping applications, including pharmaceutical analysis, functional imaging, and space exploration.

\section{\Large{Discussion}}
In this paper, we propose HYperspectral Penetrating-type Ellipsoidal ReconstructION (HYPERION) to blindly unmix the sources (i.e., the THz absorption spectra of pure substances) from transmitted THz signals, which are usually noisy and low data purity.
In HYPERION, affine fitting and simplex fitting are utilized to address the noisy and low data purity issues, respectively. In affine fitting, the noise is filtered out by projecting the data into the lower dimensional convex hull; in simplex fitting, HYPERION is encouraged to fit a regular simplex to the data convex geometry in the preconditioned space, where a relatively mild data purity is required (i.e., $\gamma>\frac{1}{\sqrt{q-1}}$). Additionally, the THz spectra with the overlapped material absorption peak in the measured THz bands, such as D-glucose and D-sucrose in 0.2 - 1.75 THz bands, can be ill-conditioned and lead to inferior performance. To address the ill-conditioned case, HYPERION transforms the transmitted THz signals into the preconditioned space, where the LJE information can easily separate the similar THz spectra. Based on the affine and simplex fittings and the transformation based on LJE information, HYPERION is capable of handling the ill-conditioned THz signals with noise and low data purity.

To evaluate the unmixing efficacy of HYPERION upon the ill-conditioned dataset with low data purity, we selected the quinary dataset without pure substances for the qualitative and quantitative analysis. In the ill-conditioned case, the THz material absorption spectra of D-glucose and D-sucrose are well-unmixed with RMSE of $2.59~\text{cm}^{-1}$ and $3.08~\text{cm}^{-1}$ compared with the ground truth spectra, respectively. In the low data purity case, even though the dataset only contains the highly mixed tablets (i.e., 5:5 for every pair of the pure substances), HYPERION still demonstrates the comparable unmixing performance of the dataset with pure substances. Note that, compared to the 1:9 mixing condition, the 5:5 mixing condition is indeed highly mixed. Additionally, for the noisy condition, we have also applied AWGN with 0.001\% - 0.1\% SD range to the quinary dataset without pure substances. In the SD range under 0.04\%, HYPERION delivers similar unmixing performance; in the SD range higher than 0.04\%, the unmixing performance is slightly inferior since the SNR of THz signals drops dramatically in the high-frequency regime. With the detailed evaluation, HYPERION reveals the great unmixing efficacy on complex mixture systems under noisy conditions.

With the support of complex mixture systems, HYPERION is capable of contributing to the applications, including pharmaceutical analysis, biomedical diagnosis \cite{sun2017recent}, and art conservation \cite{seco2013goya}. Among those applications, pharmaceutical analysis is much more challenging since it requires identification of drugs and monitoring the impurity of drugs \cite{lawrence2004applications}. In this sense, we have demonstrated how HYPERION can further help analyze drug compositions and impurity levels without prior drug recipe information. In the demonstration, a material absorption spectrum dataset, composed of the mixtures of D-lactose monohydrate, D-glucose, and L-tyrosine with distinctive material composition, is adopted to simulate the different impurity levels of drugs. Upon the dataset, THz HYPERION can precisely estimate most of the mixture composition and deliver less than 20\% inaccuracy of each composition compared with ground truth.

THz hyperspectral imaging is a great extent for HYPERION for future work since the LJE can be more accurately estimated from a large number of spectra. The task-oriented regularizers can also be designed for the spatial relation of pixels, such as spatial continuity and self-similarity. To this extent, more precise unmixed material absorption spectra and complex material mapping with a wide composition range can be well-resolved. Furthermore, the more efficient optimizer can significantly decrease the computation time, which utilizes a large number of spectra in hyperspectral images. In addition to the above future work for the more precise estimation of unmixed spectra and material mapping, the required information of tablet thickness in HYPERION can be relaxed by combining with the methods estimating the complex refractive index based on THz-TDS systems \cite{dorney2001material, duvillaret1999highly}. By relaxing the required information of tablet thickness, HYPERION can further contribute to the applications where the sample thickness cannot be measured and obtained. 

\section{\Large{Methods}}

\subsection*{L\"owner-John ellipsoid (LJE) for transmitted (penetrating-type) THz signals}
In complex mixture systems, the material absorption spectra of the overlapped material absorption peaks in the measured THz bands can be quite similar (i.e., D-sucrose and D-glucose).
It is the so-called ill-conditioned HU problem, having drawn attention from very recent machine learning literature \cite{HISUN}.
As far as we know, the most effective solution for addressing the ill-conditioned HU is based on the L\"owner-John ellipsoid (LJE) theory \cite{HISUN}, which elegantly exploits the data convex structure (i.e., the convex hull of signature vectors contains the data vectors; cf. equation \eqref{eq:cvx-stucture}).
However, almost all the HU theories (including LJE) were developed for the reflecting-type signals, probably because the need for the HU techniques mainly comes from the satellite hyperspectral remote sensing, for which the hyperspectral signals are reflected from the objects to the satellite sensors \cite{Lin2015icasspHyperCSITSP}.
To apply the LJE theory on the transmitted THz signals, our first task is to reveal the data convex structure of the THz signals.

Let $m_i(t),~i=1,\dots,n$ be the impulse response function of the $i^\text{th}$ material that the input THz signal $x(t)$ penetrates through, and let $\otimes$ denote the convolution operator.
The transmitted THz signal measured at time $t$ can then be modeled as
\[
y(t)
=x(t)\otimes m_1(t)\otimes\cdots\otimes m_n(t).
\]
By taking the Fourier transform on both sides, we then have
\[
Y(f)
=X(f)~\!M_1(f) \cdots M_n(f),
\]
where $(Y,X,M_i)$ are the Fourier transforms of $(y,x,m_i)$, respectively, and $f$ is the frequency index.
If there are $k$ frequency samples $f_1,\dots,f_k$ in the THz spectral regions, then $(Y,X,M_i)$ can be considered as $k$-dimensional vectors (e.g., $Y=[Y(f_1),\dots,Y(f_k)]^T$).
Let us define the standardized data as $Z(f)=\log\left(\frac{|Y(f)|}{|X(f)|}\right)=\sum_{i=1}^{n} {\log(|M_i(f)|)}$. 
Also, the materials $M_1,\dots,M_n$ may not be distinct, we assume that there are $q$ distinct materials in the set $\{M_1, \dots, M_n\}$, and let $N(i)$ be the set of indices corresponding to the $i^\text{th}$ distinct material in the set $\{M_1, \dots, M_n\}$. 
By defining $H_i(f)= \prod_{j \in N(i)} M_j(f)$,
we can further simplify the expression as $Z(f)=\sum_{i=1}^{q} {\log(|H_i(f)|)}$. 
Here, we neglect the energy loss between the interfaces of materials since it is relatively small compared to energy loss in the lossy medium (Supplementary Note 4). According to the derived approximation of material absorption coefficients (Supplementary Note 4), we have $\log(|H_i(f)|)=\frac{1}{2}\alpha_i(f)d_i,~i=1,\dots,q$, thereby leading to the standardized data representation
\[
Z(f)=\sum_{i=1}^{q} \frac{1}{2}\alpha_i(f)d_i,
\]
where $\alpha_i(f)$ is the material absorption coefficient of the $i^\text{th}$ distinct material at frequency $f$, and $d_i$ is the penetration depth of the $i^\text{th}$ distinct material.

Naturally, we define the hyperspectral signature of the $i^\text{th}$ material as 
\[
\bm s_i
\triangleq
[\alpha_i(f_1),\dots,\alpha_i(f_k)]^T\in\mathbb{R}^k,~i=1,\dots,q.
\]
Let $d_i^n$ be the penetration depth of the $i^\text{th}$ distinct material in the $n^\text{th}$ sample data.
It is elegant to observe that if we have the information of the value of $l_n\triangleq\frac{1}{2}(d_1^n+\dots+d_q^n)$, simply normalizing the $n^\text{th}$ standardized data $Z_n\triangleq[~\!Z_n(f_1),\dots,Z_n(f_k)~\!]^T$ by $l_n$ can reveal the desired convex structure, i.e., $\frac{Z_n}{l_n}\in\textrm{conv}\{\ensuremath{{\bm s}}_1,\dots,\ensuremath{{\bm s}}_q\}$.
If the incident angle is zero, $l_n$ is nothing but half of the thickness of the $n^\text{th}$ sample. 
Therefore, in the proposed HYPERION algorithm, the $\ell$ samples $Z_1,\dots,Z_\ell$ are allowed to have different thickness $2l_1,\dots,2l_\ell$.
Based on different thickness of samples, the normalized/standardized data $\ensuremath{{\bm x}}_n=\left[\frac{Z_n(f_1)}{l_n },\dots,\frac{Z_n(f_k)}{l_n }\right]^T$ have the desired data convex structure
\begin{equation}
\label{eq:cvx-stucture}
\ensuremath{{\bm x}}_1,\dots,\ensuremath{{\bm x}}_\ell\in\textrm{conv}\{\ensuremath{{\bm s}}_1,\dots,\ensuremath{{\bm s}}_q\}.
\end{equation}
The aim is to recover the THz signatures $\ensuremath{{\bm s}}_1,\dots,\ensuremath{{\bm s}}_q$ from the preprocessed THz transmitted signals $\ensuremath{{\bm x}}_1,\dots,\ensuremath{{\bm x}}_\ell$ for material identification.

\begin{figure}
    \centering
    \includegraphics[width=1.0\linewidth]{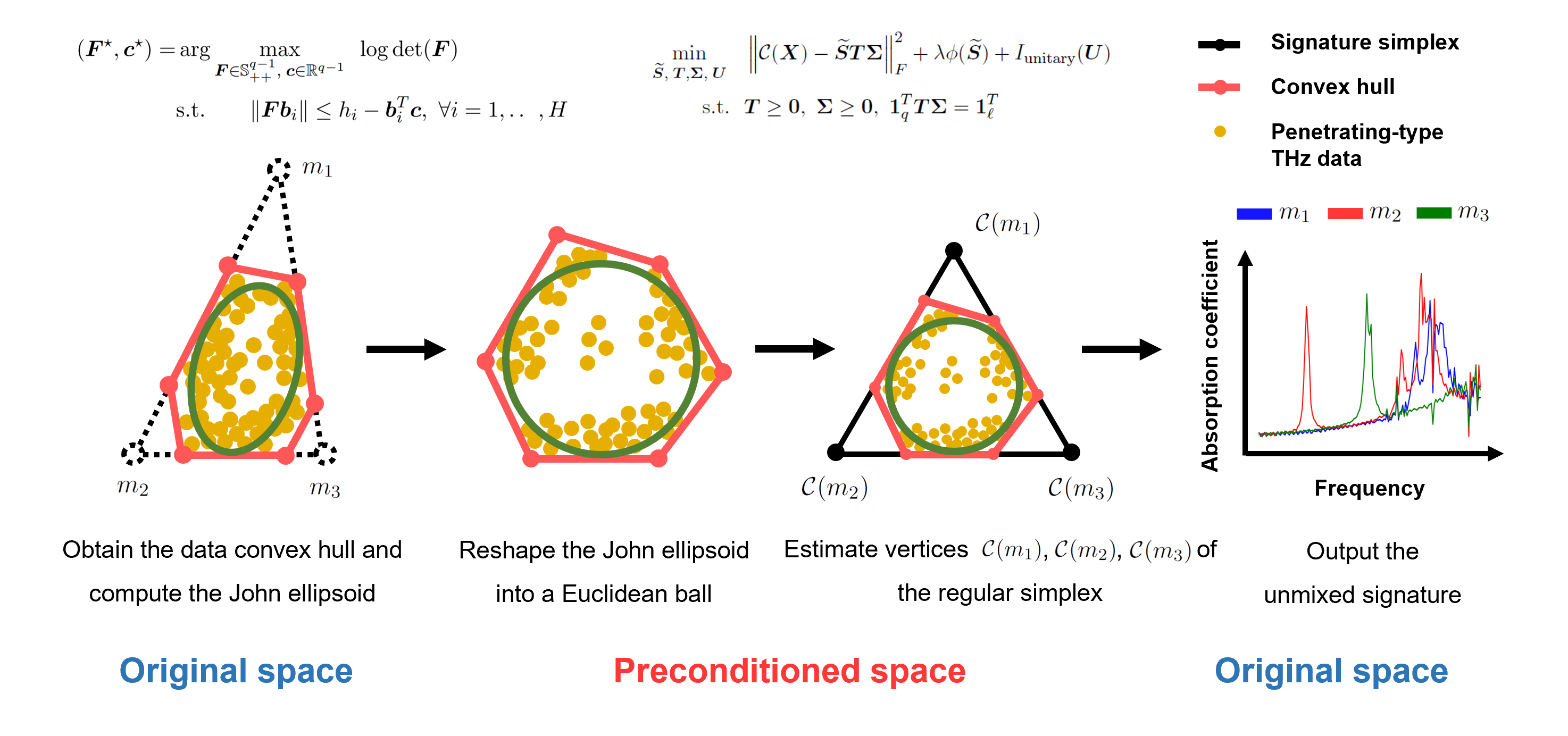}
    \captionsetup{font={small,stretch=1.25}}
    \caption{\textbf{Illustration of HYPERION.} The illustration is taken in the ternary case for comprehensive visualization. The points, $m_1$, $m_2$ and $m_3$ are the unknown THz sources of the complex mixture system. The LJE is first obtained by the convex hull of data points in the original space. The transformation between the obtained LJE and a Euclidean ball, $F$, and the center of the LJE, $\bm c$ are then utilized to transform data points to the preconditioned space for addressing the ill-conditioned case. Since the signatures of precondition operator, $\mathcal{C}(m_1)$, $\mathcal{C}(m_2)$ and $\mathcal{C}(m_3)$, should form corners of a simplex with provable regular structure in the noise-free case, HYPERION is encouraged to fit a regular simplex to the data convex geometry. To extract the unmixed signatures, the corners of the constructed simplex are transformed back to the original space by the inverse function of precondition operator.}
    \label{fig:convex}
\end{figure}

Due to the convex structure in equation \eqref{eq:cvx-stucture}, HYPERION can use the information of LJE, defined as the maximum-volume ellipsoid inscribed in the standardized/normalized penetrating-type THz signals as shown in Fig. \ref{fig:convex}, for the source separation with mild requirement of data purity (Supplementary Note 5). 

To be precise, the standardized/normalized transmitted THz data matrix $\ensuremath{{\bm X}}$ can be explicitly written as
\[
\ensuremath{{\bm X}}
\triangleq [\ensuremath{{\bm x}}_1,\dots,\ensuremath{{\bm x}}_\ell]
=\ensuremath{{\bm P}}\ensuremath{{\bm T}}\bm{\Sigma}\in\mathbb{R}^{k\times\ell},
\]
where $\ensuremath{{\bm P}}\in\mathbb{R}^{k\times q}$ with $[\ensuremath{{\bm P}}]_{ij}\triangleq \alpha_j(f_i)$, $\ensuremath{{\bm T}}\in \mathbb{R}^{q\times \ell}$ with $[\ensuremath{{\bm T}}]_{ij}\triangleq \frac{1}{2}d_i^j$, and $\bm{\Sigma}\in\mathbb{R}^{\ell\times\ell}$ is the diagonal matrix with the $i^\text{th}$ diagonal entry being $\frac{1}{l_{i}}$.
Then, computing the LJE of the THz signals $\ensuremath{{\bm X}}$ can be proven to be a convex optimization problem \cite{CVXbookCLL2016} as below:
\begin{equation}\label{prob:LJE}
	\begin{aligned}
		(\bm F^\star,\bm c^\star)=&\arg\max_{{\bm F}\in\mathbb{S}^{q-1}_{++},~\!{\bm c}\in\mathbb{R}^{q-1}}  ~ \log\det({\bm F}) \\
		&{\rm s.t.} ~~~~~ ~ \lVert\bm F \bm b_i\rVert \leq h_i-\bm b_i^T \bm c,~\forall i=1,\dots,H,
	\end{aligned}
\end{equation}
where $\mathbb{S}^{p-1}_{++}$ is the positive semidefinite (PSD) cone, the halfspace parameters $\{(\ensuremath{{\bm b}}_1,h_1),\dots,(\ensuremath{{\bm b}}_H,h_H)\}$ come from the $\cal H$-polytope representation of the transmitted THz data $\ensuremath{{\bm X}}$, i.e., $\mathcal{H}(\ensuremath{{\bm X}})=\mathcal{H}(\ensuremath{{\bm P}}\ensuremath{{\bm T}}\bm{\Sigma})\equiv\mathcal{H}(\textrm{conv}\{\bm x_1,\dots,\bm x_\ell\})=\bigcap_{i=1}^H~\!\{\bm x\mid \bm b_i^T \bm x \leq h_i \}$ \cite{barber1996quickhull}, and the optimal argument $(\bm F^\star,\bm c^\star)$ gives the desired maximum-volume ellipsoid inscribed in the THz data $\ensuremath{{\bm X}}\equiv\mathcal{H}(\ensuremath{{\bm X}})$ (i.e., LJE) as
\[
\mathcal{E}(\bm F^\star,\bm c^\star)\triangleq \{\bm F^\star\boldsymbol{\alpha}+\bm c^\star\mid\|\boldsymbol{\alpha}\|\leq 1\}\subseteq\mathbb{R}^{q-1}.
\]
Note that we have assumed w.l.o.g. that the dimension of $\textrm{conv}\{\bm x_1,\dots,\bm x_\ell\}$ is no greater than $q-1$, as implied by equation \eqref{eq:cvx-stucture}.
The solving details for equation \eqref{prob:LJE} are discussed in the Supplementary Note 7.

\subsection*{LJE $\mathcal{E}(\bm F^\star,\bm c^\star)$ for solving the transmitted THz unmixing problem}

As aforementioned, THz data is ill-conditioned.
As reported in \cite{HISUN}, the information embedded in the LJE $\mathcal{E}(\ensuremath{{\bm F}}^\star,\ensuremath{{\bm c}}^\star)$ is critical in preconditioning the data $\ensuremath{{\bm X}}$; specifically, LJE yields the preconditioned data
\[
\mathcal{C}(\ensuremath{{\bm X}})
=
(\ensuremath{{\bm F}}^\star)^\dagger\left(
\ensuremath{{\bm P}}\ensuremath{{\bm T}}\bm{\Sigma}
-\ensuremath{{\bm c}}^\star\bm 1_\ell^T
\right)
\in\mathbb{R}^{(q-1)\times\ell},
\]
whose corresponding (preconditioned) THz signatures are easily verified to be $\widetilde{\ensuremath{{\bm S}}}=
(\ensuremath{{\bm F}}^\star)^\dagger
\left(
[\ensuremath{{\bm s}}_1,\dots,\ensuremath{{\bm s}}_q]
-\ensuremath{{\bm c}}^\star\bm 1_q^T
\right)$;
note that the precondition operator $\mathcal{C}(\cdot)$ is applied columnwisely as $\mathcal{C}(\ensuremath{{\bm v}})=(\ensuremath{{\bm F}}^\star)^\dagger(\ensuremath{{\bm v}}-\ensuremath{{\bm c}}^\star)$ for any given (column) vector $\ensuremath{{\bm v}}$.
Remarkably, estimating the preconditioned THz signatures $\widetilde{\ensuremath{{\bm S}}}$ is much more friendly because, according to the LJE theory \cite[Theorem 1]{HISUN}, columns of $\widetilde{\ensuremath{{\bm S}}}$ will form a regular simplex centered at the origin whenever the data purity $\gamma>\frac{1}{\sqrt{q-1}}$.
As $\widetilde{\ensuremath{{\bm S}}}$ is obtained, the THz signatures can be simply recovered as $[\ensuremath{{\bm s}}_1,\dots,\ensuremath{{\bm s}}_q]=(\ensuremath{{\bm F}}^\star)^\dagger\widetilde{\ensuremath{{\bm S}}}+\ensuremath{{\bm c}}^\star\bm 1_q^T$.
Therefore, we can focus on estimating $\widetilde{\ensuremath{{\bm S}}}$ from the preconditioned THz data $\mathcal{C}(\ensuremath{{\bm X}})$ next.

According to the regular simplex structure, $\widetilde{\ensuremath{{\bm S}}}$ can be characterized as \cite[Equation (11)]{HISUN}
\begin{equation}\label{eq:RSCO}
\widetilde{\ensuremath{{\bm S}}}=\alpha\ensuremath{{\bm U}}^T\ensuremath{{\bm S}}_0,
\end{equation}
where $\ensuremath{{\bm S}}_0$ forms any {\it unit-volume} regular simplex centered at the origin $\bm 0_{q-1}$, with closed-form expression available in \cite[Proposition 2]{HISUN};
the unitary matrix $\ensuremath{{\bm U}}$ (i.e., $\ensuremath{{\bm U}}^T\ensuremath{{\bm U}}=\bm I_{q-1}$) and the scalar $\alpha>0$ (with closed-form expression available in \cite[Equation (12)]{HISUN}) are for rotating and scaling $\ensuremath{{\bm S}}_0$ to fit $\widetilde{\ensuremath{{\bm S}}}$.

As the transmitted THz data is not just ill-conditioned but also noisy, the regular structure may not be strictly satisfied.
Therefore, unlike \cite{HISUN}, we did not force the regular structure.
Instead, we just use such information to design a regularizer $\phi(\widetilde{\ensuremath{{\bm S}}})\triangleq\left\|\widetilde{\ensuremath{{\bm S}}}-\alpha\bm U^T\ensuremath{{\bm S}}_0\right\|_F^2$, by minimizing which we just encourage the regular structure (rather than forcing it).

To finish the design of the HU criterion for the transmitted THz data, we need to design the data-fitting term.
To this end, we need the following lemma, whose proof is given in Supplementary Note 8.
\begin{Lemma}\label{lemma:data-fit}
The precondition operator $\mathcal{C}$ satisfies the relation of $\mathcal{C}(\ensuremath{{\bm X}})=\widetilde{\ensuremath{{\bm S}}}\ensuremath{{\bm T}}\bm{\Sigma}$.\hfill$\square$
\end{Lemma}
By Lemma \ref{lemma:data-fit}, we naturally design the data fitting term as $\left\| \mathcal{C}(\ensuremath{{\bm X}})-\widetilde{\ensuremath{{\bm S}}}\ensuremath{{\bm T}}\bm{\Sigma}\right\|_F^2$, which, together with the regularizer $\phi$, yields the following HU criterion for transmitted THz signals:
\begin{align}
\label{prob:ICHU-formulation_new}
	\min_{\widetilde{\ensuremath{{\bm S}}},~\!\ensuremath{{\bm T}},\bm{\Sigma},~\!\bm U}~&~\left\| \mathcal{C}(\ensuremath{{\bm X}})-\widetilde{\ensuremath{{\bm S}}}\ensuremath{{\bm T}}\bm{\Sigma}\right\|_F^2
	+\lambda 
	\phi(\widetilde{\ensuremath{{\bm S}}})
	+I_\textrm{unitary}(\ensuremath{{\bm U}})
	\\
	\textrm{s.t.}~&~
	\ensuremath{{\bm T}}\geq\bm 0,~
	\bm{\Sigma}\geq\bm 0,~
	\bm 1_q^T\ensuremath{{\bm T}}\bm{\Sigma}=\bm 1_\ell^T,\nonumber
 \end{align}
where $\lambda:=1$ is the regularization parameter, and $I_\textrm{unitary}(\cdot)$ is the indicator function of the set of unitary matrices.
An algorithm for solving equation \eqref{prob:ICHU-formulation_new} is provided in Supplementary Note 9.
Once equation \eqref{prob:ICHU-formulation_new} is solved, the column vectors of the optimal solution $\widetilde{\ensuremath{{\bm S}}}^\star$ then serves as the estimates of the THz signatures in the preconditioned space as shown in Fig. \ref{fig:convex}. Based on the problem formulation and the solving detail, the algorithm, termed HYperspectral Penetrating-type Ellipsoidal ReconstructION (HYPERION), has been completed. Remarkably, HYPERION does not use any information about the pattern of resonant peaks of the signatures and is designed under a fully unsupervised setting.

\subsection*{\normalsize{ASOPS THz-TDS System}}

\begin{figure}
    \centering
    \includegraphics[width=1.0\linewidth]{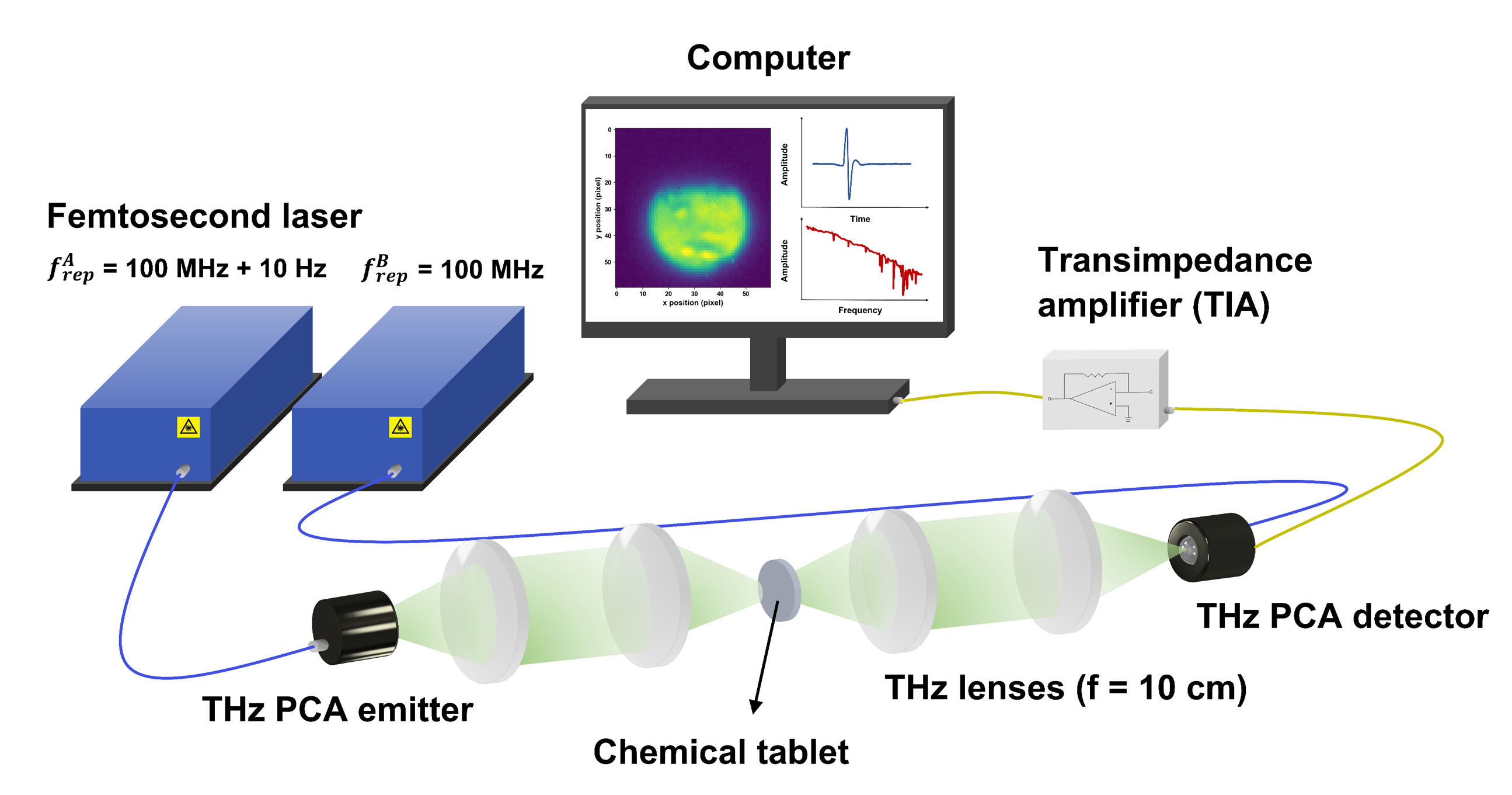}
    \captionsetup{font={small,stretch=1.25}}
    \caption{\textbf{Experimental setup.} The THz photoconductive antenna (PCA) emitter and detector are excited by two asynchronized femtosecond lasers with repetition rate ($f_\text{rep}$) of 100 MHz + 10 Hz and 100 MHz. The repetition rate difference ($\Delta f$) of two lasers is 10 Hz. The generated THz beam is focused to a spot with a diameter of 1.5 mm by THz convex lenses (focal length: 10 cm), interacting with chemical tablets and being detected by the THz photoconductive antenna detector. The detected THz electric field is converted to photocurrent and is further amplified by a transimpedance amplifier (TIA). The computer converts the amplified analog signal to the digital data and presents the measured THz signals in both time and frequency domains.}
    \label{fig:system_setup}
\end{figure}

In the asynchronized optical sampling (ASOPS) THz-TDS system (Menlo TERA ASOPS, Menlo Systems, Germany) as shown in Fig. \ref{fig:system_setup}, two asynchronized Er-doped fiber femtosecond lasers are fed into an InGaAs/InAlAs THz photoconductive antenna emitter and an LT-InGaAs/InAlAs THz detector. The power, bandwidth, and dynamic range of the THz-TDS system are up to 60 µW, less than 4.5 THz, and greater than 80 dB. 
The generated THz radiation from the THz photoconductive antenna emitter then consecutively travels through two convex THz lenses with a focal length of 10 cm. The first convex lens is used for THz beam collimation, and the second convex lens is to focus the THz radiations on the tablets. 
The diameter of the focused beam is approximately 1.5 mm of full width at half maximum (FWHM). After penetrating through tablets, the transmitted THz waves contained material absorption information in THz range, then traveling through two identical convex THz lenses to the THz photoconductive antenna detector. 
THz photoconductive antenna detector is used to retrieve the time-resolved THz electric field and convert the electric field signal to the photocurrent. The connected transimpedance amplifier (TIA) then amplifies the induced photocurrent to voltage signals.
The bandwidth and the gain of the TIA are 1.8 MHz and 10$^6$ (V/A), respectively. Each measurement contains a 100 ps time-domain trace with a 5 fs temporal resolution. 
To further increase the signal-to-noise ratio (SNR) of the system, we average 1,000 time-domain traces, which can effectively decrease the time-domain additive white Gaussian noise to 0.015\% according to the law of large numbers (Supplementary Note 10). With the setting above, the system offers a dynamic range of over 65 dB from 0.1 THz to 3 THz.

\subsection*{\normalsize{Tablet Preparation}}

In this study, all tablets were composed of D-glucose, D-lactose monohydrate, L-histidine, L-tyrosine, and D-sucrose powders. 
D-glucose, D-lactose monohydrate, and D-sucrose powders were from Merck \& Co., Inc. (Kenilworth, NJ, USA). L-histidine, L-tyrosine powders were from Acros Organics (Geel, Belgium). 
Each chemical was ground and mixed uniformly by the ball mill, ensuring that the Mie scattering is greatly reduced in the measured spectral characteristics. 
Followed by the pulverization, the chemical powders were poured into a tablet die made by a high-speed steel (HSS) for subsequent high-pressure compression. Powders were compressed by a hydraulic press (Specac, Orpington, U.K.) to form 3 mm-thick tablets (total mass: 0.3 g; thickness: $3.05\pm0.02$ mm; see Supplementary Table 6 for the details) under the pressure of 1,000 kg/cm$^2$ for 15 seconds. 
Tablets were then mounted on a 0.6 mm thick polylactic acid (PLA) plate. The copper foil is covered around the tablets to block THz signals encoding the PLA plate material information. 
Additionally, the copper foil prevents the low-frequency diffraction effect caused by the edge of the PLA plate.

\section{\Large{Supplementary Information}}

\subsection*{\large{Supplementary Note 1: Measurement Protocol}}

\begin{figure}
  \centering
  \includegraphics[width=1.0\linewidth]{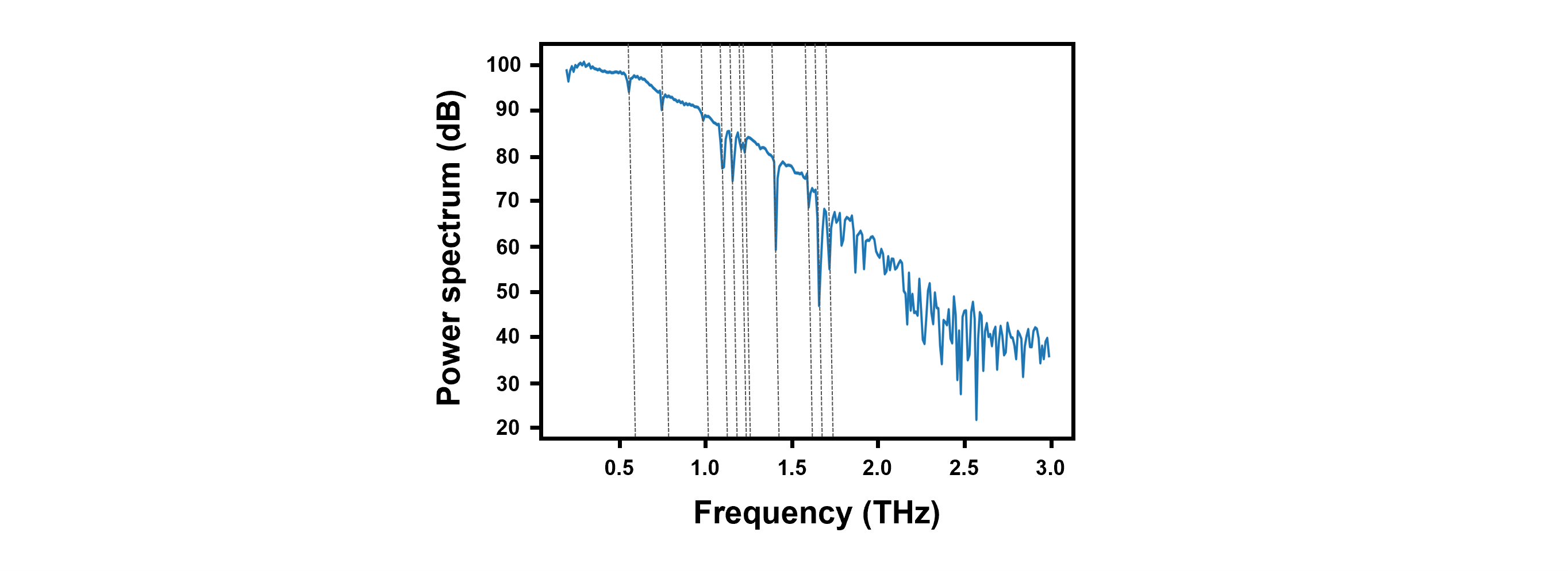}
  \captionsetup{font={small,stretch=1.25}}
  \caption{\textbf{The THz frequency-domain spectrum of air.} Water vapor absorbs THz wave at 0.56 THz, 0.75 THz, 0.99 THz, 1.10 THz, 1.16 THz, 1.21 THz, 1.23 THz, 1.41 THz, 1.60 THz, 1.66 THz, and 1.72 THz within 0.2 - 1.75 THz frequency range. The absorption lines are indicated by gray dotted lines.}
  \label{fig:ref_spectrum}
\end{figure}

Each spectrum is obtained by fast Fourier transform (FFT) of a 100 picoseconds (ps) THz time-domain trace based on an asynchronous optical sampling (ASOPS) THz-TDS system (see Supplementary Note 11). To increase the system dynamic range, 1,000 spectra of each mixture are measured and averaged, providing more than 60 dB dynamic range at 0.3 THz and 25 dB dynamic range at 1.75 THz. All the measurements are conducted at room temperature (23°C), one atmosphere (atm) pressure, and 60\% humidity. Under this condition, the dataset contains water vapor absorption lines at 0.56 THz, 0.75 THz, 0.99 THz, 1.10 THz, 1.16 THz, 1.21 THz, 1.23 THz, 1.41 THz, 1.60 THz, 1.66 THz, and 1.72 THz within 0.2 - 1.75 THz frequency range (Supplementary Fig. \ref{fig:ref_spectrum}), which contribute nearly 2.3\% of the nonlinear region in the measured frequency bands.

\subsection*{\large{Supplementary Note 2: Linear Mixing Model Validation}}

Whether the dataset follows the linear mixing model is essential since HYPERION utilizes the convex geometry of the model to unmix the material signatures. 
To validate the linear mixing model upon the dataset, principle component analysis (PCA) is one of the most suitable tools since data information can still be reserved after projecting to a lower-dimensional space \cite{pedregosa2011scikit}. To be precise, we project the non-pure mixture signatures to the optimal 2 dimensional (2D) space, which is constructed by the two eigenvectors corresponding to the two largest eigenvalues. Additionally, only non-pure mixture points are selected for the calculation of the 2D space since the assumption of pure substances is not held. After obtaining the optimal 2D plane, the data points of pure substances are further projected to the optimal 2D plane.
In the evaluation of the linear mixing model, the quinary dataset with pure substances is selected due to the material variety. 
As shown in Supplementary Fig. \ref{fig:pca_result}, the high-dimensional data points are projected onto the two-dimensional plane. Except for the extreme cases, most of the data points in every subplot form the triangle in the projected 2D space, where three points lie at the vertex and three points fall at the center of the side. 
The misaligned data points are caused by the inaccurate estimation of the two-dimensional plane, which is caused by the measurement noise and nonlinear light-matter interaction. Based on the projected data points, the measured data points are highly aligned with the linear mixing model. Additionally, the triangles involving D-glucose and D-lactose are narrower since the signatures of the two chemicals are the ill-conditioned case.
It is worthwhile to mention that the alignment accuracy between the model and measured data may be further enhanced by modeling the nonlinear matter-matter interaction into the linear mixing model.

\begin{figure}
  \centering
  \includegraphics[width=1.0\linewidth]{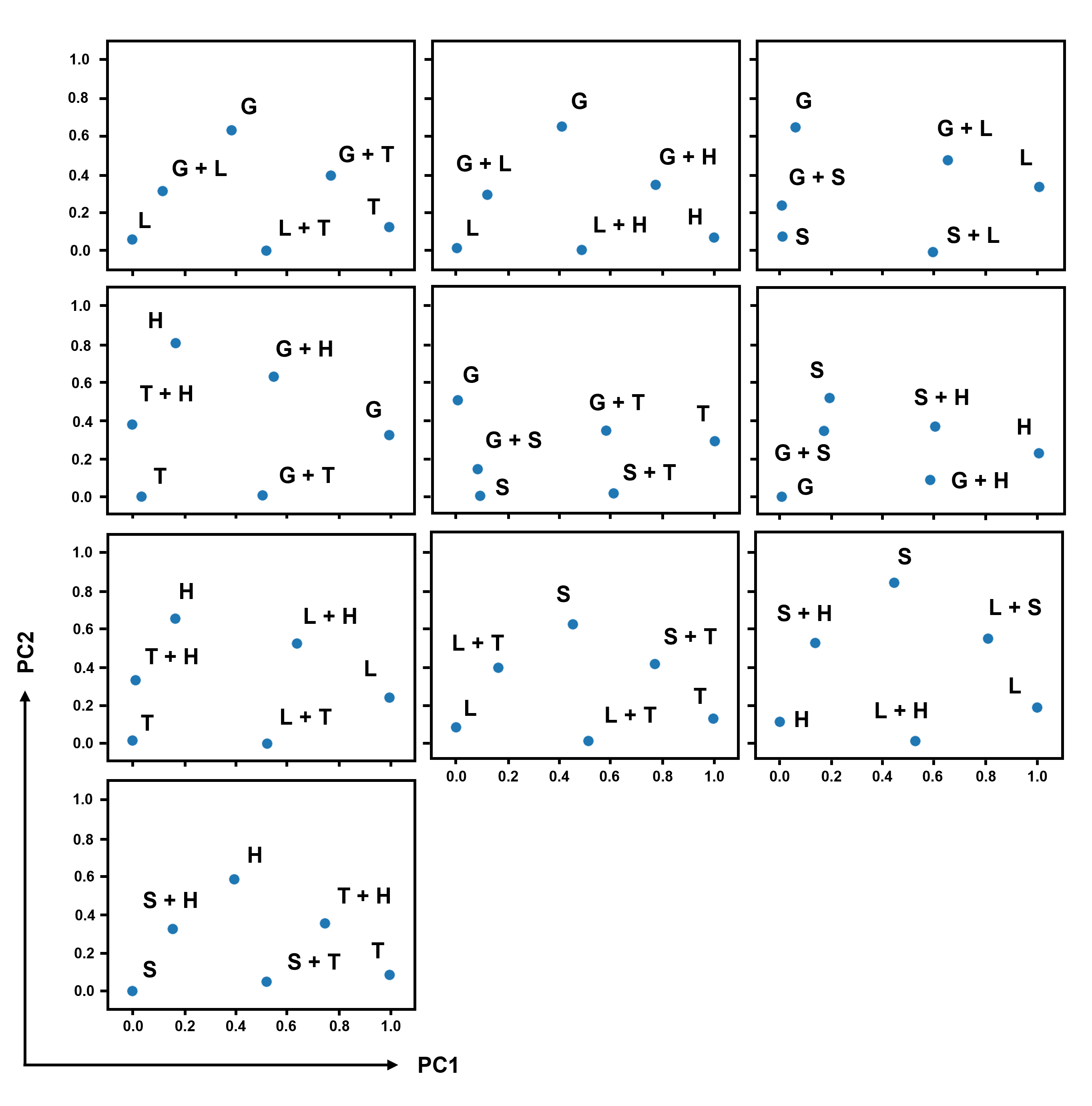}
  \captionsetup{font={small,stretch=1.25}}
  \caption{\textbf{Prinpical component analysis (PCA) projection of the quinary dataset}. Data points from the quinary dataset with pure substances are projected on a two-dimensional projection plane by PCA. Five pure substances are D-glucose, D-lactose monohydrate, D-sucrose, L-tyrosine, and L-histidine, which are abbreviated as G, L, S, T, and H.}
  \label{fig:pca_result}
\end{figure}

\subsection*{\large{Supplementary Note 3: Spectral Angle Mapper (SAM) and Root Mean Square Error (RMSE)}}

Spectral angle mapper (SAM) is typically used to evaluate the shape similarity between two spectra. The definition is as a following equation:
\begin{equation}
    \label{eq:error_sae}
    \theta(s_1,s_2)=\cos^{-1}\left (\frac{s_1\cdot s_2}{\left\|s_1 \right\|_2\cdot \left\|s_2 \right\|_2} \right), 
\end{equation}
where $s_1$ and $s_2$ are spectra.\\
Root mean square error (RMSE) does not only focus on the shape similarity but the band-wise differences. RMSE is defined as a following equation:
\begin{equation}
    \label{eq:error_rmse}
    d(s_1,s_2)=\sqrt{\frac{\left\|s_1-s_2 \right\|_2^2}{n}}, 
\end{equation}
where $n$ is the dimension of the spectrum.

\subsection*{\large{Supplementary Note 4: Data Preprocessing}}

The THz-TDS system provides a time-domain waveform that profiles the interactions between THz radiation and measured tablets. The reference waveform $S_\text{ref}(t)$ and the tablet waveform $S_\text{tablet}(t)$, where $t$ is the time index, refer to the measured waveform without and with a tablet, respectively. 
After obtaining reference waveform $S_\text{ref}(t)$ and tablet waveform $S_\text{tablet}(t)$, the fast Fourier transform (FFT) was applied to convert time-domain waveforms to the frequency-domain reference spectrum $S_\text{ref}(\omega)$ and tablet spectrum $S_\text{tablet}(\omega)$, where $\omega$ is the angular frequency. 
To further extract the material absorption spectrum ($\alpha(\omega)$) of the tablet, the thickness of the tablet is an important parameter for the derivation. 
The thickness of the tablet is measured by the electric ruler, which provides a 0.1 mm resolution. Compared to the thickness of the tablet (i.e., ~3 mm), this resolution of 0.1 mm introduces approximately 3\% inaccuracy in the measurement.
The material absorption spectra of the tablets are derived from the Fresnel equations and the THz wave propagation model based on homogeneous and planar materials. Furthermore, according to this derivation, the thickness measurement error of 3\% will only introduce approximately 3\% error in material absorption spectra. Since the particle size of the tablet powder is vastly smaller than the wavelength of the THz waves (Supplementary Note 16)
Additionally, the thickness of the tablet is significantly larger than the effective wavelength of the THz signal in the 0.2 - 1.75 THz frequency range such that the Fabry-Pérot effect \cite{dorney2001material} does not introduce drastic damping to the measured spectrum. 

\begin{figure}
  \centering
  \includegraphics[width=1.0\linewidth]{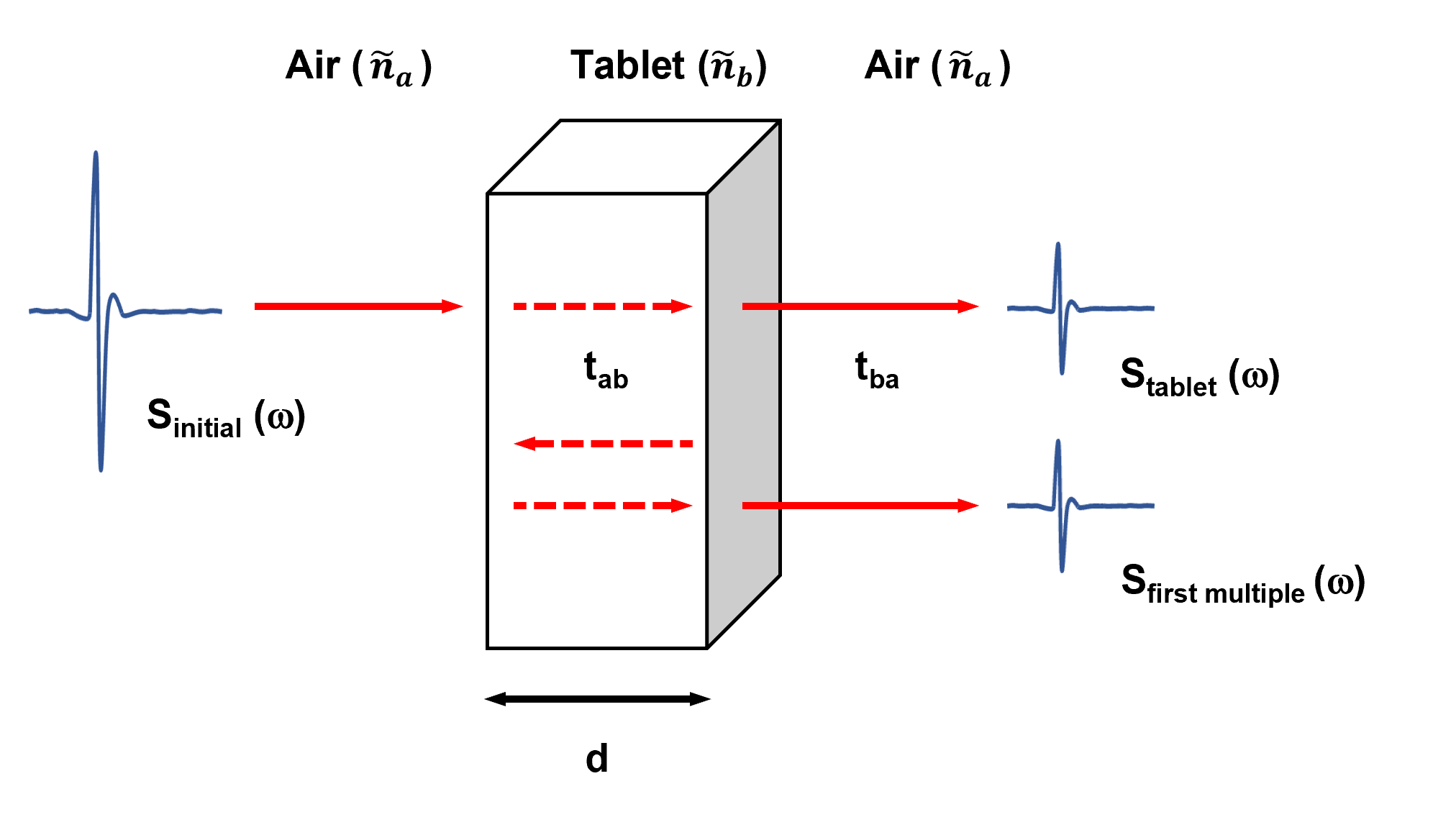} 
  \captionsetup{font={small,stretch=1.25}}
  \caption{\textbf{Illustration of THz wave traveling through a material.} A THz wave transmission and reflection pathways through a planar, homogeneous material with a thickness $d$. The THz signals shown in this figure are represented in time-domain.}
  \label{fig:propagation_model}
\end{figure}

Under this condition, we do not include the scattering influence and the Fabry-Pérot effect in this model. The Fresnel equations describe the transmission and the reflection of a THz wave traveling through an interface, which are based on the complex refractive index of the material, $\tilde{n}(\omega)=n(\omega)-j\kappa(\omega)$, where $n(\omega)$ is the real refractive index and $\kappa(\omega)$ represents the extinction coefficient which is proportional to the material absorption coefficient, $\kappa(\omega)=\left[\alpha(\omega)\cdot\lambda)\right]/4 \pi$, where $\lambda$ is the wavelength. 
The material absorption coefficients describe the power loss when the THz waves travel through the material. Considering a typical incident THz wave propagates a material with thickness $d$ (Supplementary Fig. \ref{fig:propagation_model}), the Fresnel equation \cite{dorney2001material} at an interface can be represented as
\begin{equation}
t_{ab}(\omega)=\frac{2\tilde{n}_a(\omega)}{\tilde{n}_a(\omega)+\tilde{n}_b(\omega)}, \nonumber
\end{equation}
\begin{equation}
r_{ab}(\omega)=\frac{\tilde{n}_a(\omega)-\tilde{n}_b(\omega)}{\tilde{n}_a(\omega)+\tilde{n}_b(\omega)}, \nonumber
\end{equation}
where $t_{ab}(\omega)$ is the transmission coefficient of a THz wave from the region $a$ (which is air in our experiment) to region $b$ (which is the tablet in our experiment) and $r_{ab}(\omega)$ is the reflection coefficient of a THz wave at the $a$-$b$ interface. 
Additionally, $\tilde{n}_a(\omega)$ and $\tilde{n}_b(\omega)$) represent the complex refractive index of region $a$ and $b$, respectively. The THz wave propagation constant, $p_b(\omega,d)$, is governed by
\begin{equation}
p_b(\omega,d)=\exp\left[\frac{-j\tilde{n}_b(\omega)\omega d}{c}\right], \nonumber
\end{equation}
where $c$ is the speed of light. Consequently, the equation for the received THz tablet signal in the frequency domain can be represented as
\begin{equation}
\begin{aligned}
    S_\text{tablet}(\omega) 
    & = S_\text{ref}(\omega)\cdot t_{ab}\cdot t_{ba}\cdot p_b(\omega,d) \nonumber \quad \textbf{(The Fresnel loss is included in $t_{ab}$ and $t_{ba}$.)}\\
    & = S_\text{ref}(\omega)\cdot t_{ab}\cdot t_{ba}\cdot \exp\left[\frac{-j\tilde{n}_b(\omega)\omega d}{c}\right] \nonumber \\
    & = S_\text{ref}(\omega)\cdot t_{ab}\cdot t_{ba}\cdot \exp\left\{\frac{-j[n_b(\omega)-j\kappa_b(\omega)]\omega d}{c}\right\} \nonumber \\
    & = S_\text{ref}(\omega)\cdot t_{ab}\cdot t_{ba}\cdot \exp\left[\frac{\kappa_b(\omega)\omega d}{c}\right]\cdot \exp\left[\frac{-jn_b(\omega)\omega d}{c}\right].
    \label{eq:sample_spectrum}
\end{aligned}
\end{equation}
The reference signal normalizes the tablet signal, and its amplitude can be obtained from equation (\ref{eq:sample_spectrum})
\begin{equation}
    \left|\frac{S_\text{tablet}(\omega)}{S_\text{ref}(\omega)}\right|=\left|t_{ab}\cdot t_{ba}\cdot \exp\left[ \frac{\kappa_b(\omega)\omega d}{c}\right] \right| \triangleq\left|H(\omega)\right|.
    \label{eq:transfer_fn}
\end{equation}
The extinction coefficient can be derived by taking the natural logarithm operation of equation (\ref{eq:transfer_fn})
\begin{equation}
    \ln(\left|H(\omega) \right|)=\frac{\kappa_b(\omega)\omega d}{c}+\ln(\left|t_{ab}\cdot t_{ba} \right|), \nonumber
\end{equation}
\begin{equation}
    \kappa_b(\omega)=\frac{c\left[\ln(\left|H(\omega) \right|)-\ln(\left|t_{ab}\cdot t_{ba} \right|) \right]}{\omega d}. \nonumber
\end{equation}
Therefore, the material absorption coefficient can be represented as
\begin{equation}
    \label{eq:alpha}
    \alpha^{'}(\omega)=\frac{4\pi\kappa(\omega)}{\lambda}=\frac{2\cdot\left[\ln(\left|H(\omega) \right|)-\ln(\left|t_{ab}\cdot t_{ba} \right|) \right]}{d}.
\end{equation}
Based on the derivation, to acquire the material absorption spectrum, we have first to calculate the transmission coefficients, $t_{ab}$ and $t_{ba}$. However, the interface power loss is relatively small since the thickness of the sample is greatly larger than the THz wavelength, and the tablet surface is smooth and flat compared to the THz wavelength. This relatively small interface power loss results in the high transmission coefficients of $t_{ab}$ and $t_{ba}$. Thus, we can neglected the second term of the right-hand side in equation (\ref{eq:alpha}) to form the equation \eqref{eq:alpha_approximation} since the multiplication of $t_{ab}$ and $t_{ba}$ approaches to $1$. 
\begin{equation} 
    \label{eq:alpha_approximation}
    \alpha^{'}(\omega)=\frac{2\cdot \ln(\left|H(\omega) \right|)}{d}=\frac{2\cdot \ln\left(\left|\frac{S_\text{tablet}(\omega)}{S_\text{ref}(\omega)}\right| \right)}{d}.
\end{equation}
To further match the notation in blind source separation fields, the equation (\ref{eq:alpha_approximation}) can be expressed in ordinary frequency, $f$, by the equation of $\omega=2\pi f$:
\begin{equation} 
    \label{eq:alpha_ordinary}
    \alpha^{'}(\omega)=\alpha^{'}(2\pi f)\triangleq\alpha(f). 
\end{equation}

\subsection*{\large{Supplementary Note 5: Mild algorithmic requirement}}

The aforementioned problem (i.e., recovering $\ensuremath{{\bm s}}_1,\dots,\ensuremath{{\bm s}}_q$ from their mixtures $\ensuremath{{\bm x}}_1,\dots,\ensuremath{{\bm x}}_\ell$) has been studied in the machine learning literature; for example, in the successive projection algorithm (SPA) \cite{SPAarora2012practical}, the problem is solved under one additional assumption called separability assumption, i.e.,
\[
\ensuremath{{\bm s}}_1,\dots,\ensuremath{{\bm s}}_q\in\{\ensuremath{{\bm x}}_1,\dots,\ensuremath{{\bm x}}_\ell\},
\]
meaning that, for each material, there is a pure observation solely composed of such material.
Under the separability assumption, one can show that $\textrm{conv}\{\ensuremath{{\bm x}}_1,\dots,\ensuremath{{\bm x}}_\ell\}=\textrm{conv}\{\ensuremath{{\bm s}}_1,\dots,\ensuremath{{\bm s}}_q\}$, and hence identifying the signatures is nothing but seeking the vertices of the observable geometry $\textrm{conv}\{\ensuremath{{\bm x}}_1,\dots,\ensuremath{{\bm x}}_\ell\}$, which is polynomial-time solvable.
Remarkably, by following the definition of data purity $\gamma\in(0,1]$, defined in \cite{lin2014identifiability}, the separability assumption holds if, and only if, $\gamma=1$ (the highest data purity).
So, the separability assumption is quite restricted, and such assumption is often violated in practical hyperspectral applications \cite{lin2014identifiability}.

In fact, the separability assumption may not hold in our THz application, as a tablet typically involves highly mixed chemicals, implying a low data purity.
In recent mathematics literature, the LJE theory has been proven to work well even with very low data purity; specifically, the signatures can be perfectly recovered under a much milder assumption $\gamma>\frac{1}{\sqrt{q-1}}$ \cite{lin2018maximum}.
Also, as aforementioned, the LJE criterion is theoretically and experimentally proven to be robust against the ill-conditioning of the signatures, hence quite suitable for our THz applications.
Therefore, in this paper, the convex LJE criterion is applied to the transmitted (penetrating-type) THz signals for the first time, to be detailed later.

We remark that there is other non-convex criterion that can also perform HU under the mild requirement of $\gamma>\frac{1}{\sqrt{q-1}}$ \cite{Craig1994,lin2014identifiability}, but such criterion cannot achieve the theoretical bound $\frac{1}{\sqrt{q-1}}$ in practice due to its non-convexity \cite{lin2018maximum}.
By contrast, the LJE criterion has the same theoretical bound, and in the meanwhile, it is experimentally shown to be able to achieve the bound due to its convexity nature \cite{lin2018maximum}.

\subsection*{\large{Supplementary Note 6: Implementation of THz HYPERION}}

To restore the chemical compositions, the equation \eqref{eq:supp_application} should be solved for every tablet.
\begin{equation}
\begin{aligned} \label{eq:supp_application}
	\min_{\bm{r}_i\in\mathbb{R}^{3}}~&~\left\| \bm{x}_i - \bm{A}\bm{r}_i\right\|_1
	\\
	\textrm{s.t.}~&~
	\ \bm{r}_i \succeq 0,
	\ \bm{1^T}\bm{r}_i = 1. 
\end{aligned}
\end{equation}
Since the equation \eqref{eq:supp_application} is convex, it can be solved by the solver \texttt{CVX} \cite{grant2008cvx} as the following pseudo-code:

\begin{align*}
		&{\rm cvx\underline{~~}begin}\\
		&~~~~~~{\rm variable~ r(3)~ nonnegative;}\\
		&~~~~~~{\rm minimize(~ norm( x-A*r, 1 ) ~);}\\
		&~~~~~~{\rm subject~ to} \\
		&~~~~~~~~~~~~{\rm sum( r)==1;}\\
		&{\rm  cvx\underline{~~}end}
\end{align*}

\subsection*{\large{Supplementary Note 7: How the LJE problem \eqref{prob:LJE_supp} can be solved in practice?}}

One of the core in HYPERION is the LJE problem: 
\begin{equation}\label{prob:LJE_supp}
	\begin{aligned}
		(\bm F^\star,\bm c^\star)=&\arg\max_{{\bm F}\in\mathbb{S}^{q-1}_{++},~\!{\bm c}\in\mathbb{R}^{q-1}}  ~ \log\det({\bm F}) \\
		&{\rm s.t.} ~~~~~ ~ \lVert\bm F \bm b_i\rVert \leq h_i-\bm b_i^T \bm c,~\forall i=1,\dots,H,
	\end{aligned}
\end{equation}
where $\mathbb{S}^{p-1}_{++}$ is the positive semidefinite (PSD) cone, the halfspace parameters $\{(\ensuremath{{\bm b}}_1,h_1),\dots,(\ensuremath{{\bm b}}_H,h_H)\}$ come from the $\cal H$-polytope representation of the data $\ensuremath{{\bm X}}$ and the optimal argument $(\bm F^\star,\bm c^\star)$ gives the desired maximum-volume ellipsoid inscribed in the penetrating-type THz data.

We remark that there is an algorithm \cite{HISUN} that can be adapted to solve equation \eqref{prob:LJE_supp} exactly, but it is mainly designed to handle the case with million-scale constraints $H$ induced by NASA's benchmark hyperspectral data \cite{green1998imaging}.
In typical THz applications, such as measuring the chemicals in a given tablet, we just have a few samples, and hence we do not need to use such a sophisticated method.
Instead, the general-purpose convex software CVX \cite{grant2008cvx} is effective and fast when the problem size is not too large, and hence is suitable to be adopted to solve equation \eqref{prob:LJE_supp}.

To adopt CVX, we need to implement the constraints and the objective function of equation \eqref{prob:LJE_supp}.
The constraints of equation \eqref{prob:LJE_supp} are the well-known Lorentz cone constraints ``$\lVert\bm F \bm b_i\rVert \leq h_i-\bm b_i^T \bm c$", theoretically equivalent to require the LJE to lie within the penetrating-type data convex hull $\mathcal{H}(\ensuremath{{\bm P}}\ensuremath{{\bm T}}\bm{\Sigma})\equiv\mathcal{H}(\textrm{conv}\{\bm x_1,\dots,\bm x_\ell\})$.
As for the objective function, for PSD matrix $\ensuremath{{\bm F}}$, we observe that
\[
\log\det(\ensuremath{{\bm F}})
\propto
\det(\ensuremath{{\bm F}})
\propto
(\det(\ensuremath{{\bm F}}))^{\frac{1}{q-1}}.
\]
So, the objective function can be implemented by the command ``${\rm det\underline{~~}rootn( \cdot )}$" supported by the \texttt{CVX} package.
Solving the LJE problem by the general-purpose convex software \texttt{CVX} normally reuqires some prior knowledge to transform the problem into the supporting functions. To this end, we provide the pseudo-code for the better understanding. The pseudo-code using \texttt{CVX} is as following:
\begin{align*}
		&{\rm cvx\underline{~~}begin}\\
		&~~~~~~{\rm variable~ F(q-1,q-1)~ symmetric;}\\
		&~~~~~~{\rm variable~ c(q-1);}\\
		&~~~~~~{\rm maximize(~ det\underline{~~}rootn( F ) ~);}\\
		&~~~~~~{\rm subject~ to} \\
		&~~~~~~~~~~~~{\rm for ~i = 1 : H}\\
		&~~~~~~~~~~~~{\rm norm(~ F*b(i,:),~2 ~) + b(i,:)'*c <= h(i);}\\ &~~~~~~~~~~~~{\rm end}\\
		&{\rm  cvx\underline{~~}end}
\end{align*}
%
It shows that \texttt{CVX} can handle a non-standard form of convex programming.

\subsection*{\large{Supplementary Note 8: Proof of Lemma 1.\hfill$\square$}}

Following the definition of precondition operator $\mathcal{C}$, we have
\begin{align*}
\mathcal{C}(\ensuremath{{\bm X}})
&\triangleq 
(\ensuremath{{\bm F}}^\star)^\dagger\left(
\ensuremath{{\bm P}}\ensuremath{{\bm T}}\bm{\Sigma}
-\ensuremath{{\bm c}}^\star\bm 1_\ell^T
\right)
\\
&=
(\ensuremath{{\bm F}}^\star)^\dagger\ensuremath{{\bm P}}\ensuremath{{\bm T}}\bm{\Sigma}
-
(\ensuremath{{\bm F}}^\star)^\dagger\ensuremath{{\bm c}}^\star\bm 1_\ell^T
\\
&=
(\ensuremath{{\bm F}}^\star)^\dagger\ensuremath{{\bm P}}\ensuremath{{\bm T}}\bm{\Sigma}
-
(\ensuremath{{\bm F}}^\star)^\dagger\ensuremath{{\bm c}}^\star\bm 1_q^T\ensuremath{{\bm T}}\bm{\Sigma}
\\
&=
(\ensuremath{{\bm F}}^\star)^\dagger[\ensuremath{{\bm s}}_1,\dots,\ensuremath{{\bm s}}_q]\ensuremath{{\bm T}}\bm{\Sigma}
-
(\ensuremath{{\bm F}}^\star)^\dagger\ensuremath{{\bm c}}^\star\bm 1_q^T\ensuremath{{\bm T}}\bm{\Sigma}
\\
&=\left[(\ensuremath{{\bm F}}^\star)^\dagger\left([\ensuremath{{\bm s}}_1,\dots,\ensuremath{{\bm s}}_q]-\ensuremath{{\bm c}}^\star\bm 1_q^T\right)\right]\ensuremath{{\bm T}}\bm{\Sigma}
\\
&=\widetilde{\ensuremath{{\bm S}}}\ensuremath{{\bm T}}\bm{\Sigma},
\end{align*}
where we have used the observation that each column of the matrix $\ensuremath{{\bm T}}\bm{\Sigma}$ actually provides a set of convex combination coefficients \cite{CVXbookCLL2016}, implying the equation ``$\bm 1_q^T\ensuremath{{\bm T}}\bm{\Sigma}=\bm 1_\ell^T$".
\hfill$\blacksquare$

\bigskip

\subsection*{\large{Supplementary Note 9: Algorithm details for solving the HYPERION}}

As there are four variables in equation \eqref{prob:ICHU-formulation_new_supp}, 
\begin{equation}
\begin{aligned}
\label{prob:ICHU-formulation_new_supp}
	\min_{\widetilde{\ensuremath{{\bm S}}},~\!\ensuremath{{\bm T}},\bm{\Sigma},~\!\bm U}~&~\left\| \mathcal{C}(\ensuremath{{\bm X}})-\widetilde{\ensuremath{{\bm S}}}\ensuremath{{\bm T}}\bm{\Sigma}\right\|_F^2
	+\lambda 
	\phi(\widetilde{\ensuremath{{\bm S}}})
	+I_\textrm{unitary}(\ensuremath{{\bm U}})
	\\
	\textrm{s.t.}~&~
	\ensuremath{{\bm T}}\geq\bm 0,~
	\bm{\Sigma}\geq\bm 0,~
	\bm 1_q^T\ensuremath{{\bm T}}\bm{\Sigma}=\bm 1_\ell^T,
\end{aligned}
\end{equation}
solving it directly using the alternating optimization may not be effective.
Considering that what we are really interested in is the THz signatures $\widetilde{\ensuremath{{\bm S}}}$, we do not need to separately optimize all the variables, allowing us to introduce the trick of changes of variables to reduce the number of block variables.
Specifically, by letting $\widetilde{\ensuremath{{\bm T}}}=\ensuremath{{\bm T}}\bm{\Sigma}$, equation \eqref{prob:ICHU-formulation_new_supp} can be reformulated as
\begin{equation}
\begin{aligned}
\label{prob:cv}
	\min_{\widetilde{\ensuremath{{\bm S}}},~\!\widetilde{\ensuremath{{\bm T}}},~\!\bm U}~&~\left\| \mathcal{C}(\ensuremath{{\bm X}})-\widetilde{\ensuremath{{\bm S}}}\widetilde{\ensuremath{{\bm T}}}\right\|_F^2
	+\lambda 
	\phi(\widetilde{\ensuremath{{\bm S}}})
	\\
	\textrm{s.t.}~&~
	{ 
	\widetilde{\ensuremath{{\bm T}}}\geq\bm 0,~
	\bm 1_q^T\widetilde{\ensuremath{{\bm T}}}=\bm 1_\ell^T,}~\bm U^T\bm U=\bm I_{q-1}.
\end{aligned}
\end{equation}
After reducing the number of block variables, it is natural to solve equation \eqref{prob:cv} using alternating optimization that alternatively updates $\widetilde{\ensuremath{{\bm S}}},~\!\widetilde{\ensuremath{{\bm T}}},~\!\bm U$, as detailed next.

The subproblem for solving $\widetilde{\ensuremath{{\bm S}}}$ (with $\widetilde{\ensuremath{{\bm T}}},~\!\bm U$ fixed) can be explicitly written as
\begin{equation}
\begin{aligned}
\widetilde{\ensuremath{{\bm S}}}^\star
=\arg\min_{\widetilde{\ensuremath{{\bm S}}}}
	\left\|\mathcal{C}(\ensuremath{{\bm X}})-\widetilde{\ensuremath{{\bm S}}}\widetilde{\ensuremath{{\bm T}}}\right\|_F^2
	+\lambda
	\left\|\widetilde{\ensuremath{{\bm S}}}-\alpha\ensuremath{{\bm U}}^T\ensuremath{{\bm S}}_0\right\|_F^2.\label{S_update}
\end{aligned}
\end{equation}
By defining $\widetilde{\bm m}\triangleq\textrm{vec}(\widetilde{\ensuremath{{\bm S}}})$, $\ensuremath{{\bm v}}\triangleq[\textrm{vec}(\mathcal{C}(\ensuremath{{\bm X}}))^T,~\sqrt{\lambda}(\textrm{vec}(\alpha\ensuremath{{\bm U}}^T\ensuremath{{\bm S}}_0 ) )^T ]^T$ and $\widetilde{\ensuremath{{\bm P}}}\triangleq[\widetilde{\ensuremath{{\bm T}}}\otimes\bm I_{q-1},~\sqrt{\lambda}\bm I_{q(q-1)} ]^T$, equation \eqref{S_update} can be simplified as
$\textrm{vec}(\widetilde{\ensuremath{{\bm S}}}^\star)\triangleq\widetilde{\bm m}^\star 
	=\arg\min_{\widetilde{\bm m}}~\|\widetilde{\ensuremath{{\bm P}}}\widetilde{\bm m}-\ensuremath{{\bm v}}\|_2^2$,
which closed-form solution can be derived as
\begin{align*}
	\widetilde{\bm m}^\star
	&=(\widetilde{\ensuremath{{\bm P}}}^T\widetilde{\ensuremath{{\bm P}}})^{-1}(\widetilde{\ensuremath{{\bm P}}}^T\ensuremath{{\bm v}})
	\\
	&=\{[\widetilde{\ensuremath{{\bm T}}}\widetilde{\ensuremath{{\bm T}}}^T+\lambda\bm I_q]^{-1}\otimes\bm I_{q-1}\}
	\textrm{vec}(\mathcal{C}(\ensuremath{{\bm X}})\widetilde{\ensuremath{{\bm T}}}^T+\lambda\alpha\ensuremath{{\bm U}}^T\ensuremath{{\bm S}}_0)
	\\
	&=\textrm{vec}\left(
	\bm I_{q-1}~\!
	[\mathcal{C}(\ensuremath{{\bm X}})\widetilde{\ensuremath{{\bm T}}}^T+\lambda\alpha\ensuremath{{\bm U}}^T\ensuremath{{\bm S}}_0]~\!
	[\widetilde{\ensuremath{{\bm T}}}\widetilde{\ensuremath{{\bm T}}}^T+\lambda\bm I_q]^{-T}
	\right),
\end{align*}
where $\otimes$ and $\textrm{vec}(\cdot)$ denote Kronecker product and vectorization operator, respectively.
Therefore, we have $\widetilde{\ensuremath{{\bm S}}}^\star =[\mathcal{C}(\ensuremath{{\bm X}})\widetilde{\ensuremath{{\bm T}}}^T+\lambda\alpha\ensuremath{{\bm U}}^T\ensuremath{{\bm S}}_0] [\widetilde{\ensuremath{{\bm T}}}\widetilde{\ensuremath{{\bm T}}}^T+\lambda\bm I_q]^{-T}$.

Next, note that the subproblem for solving $\widetilde{\ensuremath{{\bm T}}}$ (with $\widetilde{\ensuremath{{\bm S}}},~\!\bm U$ fixed) is \begin{equation}
\begin{aligned}
	\widetilde{\ensuremath{{\bm T}}}^\star
	=
	\arg\min_{\widetilde{\ensuremath{{\bm T}}}\geq\bm 0,~
		\bm 1_q^T\widetilde{\ensuremath{{\bm T}}}=\bm 1_\ell^T}~&~\left\| \mathcal{C}(\ensuremath{{\bm X}})-\widetilde{\ensuremath{{\bm S}}}\widetilde{\ensuremath{{\bm T}}}\right\|_F^2, \nonumber
\end{aligned}
\end{equation}
which can be easily solved as the well-known fully-constrained least-squares problem \cite{SUnSAL,Heinz2001}.
Finally, the subproblem for solving $\ensuremath{{\bm U}}$ (with $\widetilde{\ensuremath{{\bm S}}},~\!\widetilde{\ensuremath{{\bm T}}}$ fixed) is a non-convex optimization problem
\begin{equation}
\begin{aligned} \label{prob:U}
	\ensuremath{{\bm U}}^\star
	=\arg\min_{\bm U^T\bm U=\bm I_{q-1}}
	\left\|\widetilde{\ensuremath{{\bm S}}}-\alpha\ensuremath{{\bm U}}^T\ensuremath{{\bm S}}_0\right\|_F^2,
\end{aligned}
\end{equation}
but can be elegantly solved using the following inequalities for square matrices $\ensuremath{{\bm A}},\ensuremath{{\bm B}}$, i.e.,
$\textrm{trace}(\ensuremath{{\bm A}})\leq \|\ensuremath{{\bm A}}\|_{*}=\|\ensuremath{{\bm A}}\|_{S^1}$ and $\|\ensuremath{{\bm A}}\ensuremath{{\bm B}}\|_{S^1}\leq\|\ensuremath{{\bm A}}\|_{S^p}\|\ensuremath{{\bm B}}\|_{S^q}$,
where $\frac{1}{p}+\frac{1}{q}=1$ ($1\leq p,q\leq\infty$), $\|\cdot\|_{*}$ denotes nuclear norm, and $\|\cdot\|_{S^p}$ denotes Schatten $p$-norm.
Since $\ensuremath{{\bm U}}$ is unitary, the objective function of equation \eqref{prob:U} can be simplified as $\textrm{trace}(\ensuremath{{\bm U}}\widetilde{\ensuremath{{\bm S}}}\ensuremath{{\bm S}}_0^T)$, whose upper bound is derivable using the inequalities as $\textrm{trace}(\ensuremath{{\bm U}}\widetilde{\ensuremath{{\bm S}}}\ensuremath{{\bm S}}_0^T)
	\leq
	\|\ensuremath{{\bm U}}\widetilde{\ensuremath{{\bm S}}}\ensuremath{{\bm S}}_0^T\|_{S^1}
	\leq
	\|\ensuremath{{\bm U}}\|_{S^\infty}
	\|\widetilde{\ensuremath{{\bm S}}}\ensuremath{{\bm S}}_0^T\|_{S^1}
	=
	\|\widetilde{\ensuremath{{\bm S}}}\ensuremath{{\bm S}}_0^T\|_{S^1}$,
and is achievable by 
$\ensuremath{{\bm U}}^\star=\ensuremath{{\bm D}}_1\ensuremath{{\bm D}}_2^T$, where $\ensuremath{{\bm D}}_1,\ensuremath{{\bm D}}_2$ are obtained from the singular value decomposition of $\widetilde{\ensuremath{{\bm S}}}\ensuremath{{\bm S}}_0^T=\ensuremath{{\bm D}}_2\bm{\Sigma}'\ensuremath{{\bm D}}_1^T$.

The algorithm, termed HYperspectral Penetrating-type Ellipsoidal ReconstructION (HYPERION), has been completed.
Remarkably, HYPERION does not use any information about the pattern of resonant peaks and is designed under a fully \textit{blind}   setting.

\subsection*{\large{Supplementary Note 10: Noise Modeling of the System}}

The noise in our THz-TDS system corresponds to additive white Gaussian noise (AWGN), whose standard deviation in the single time-domain trace is approximately 0.48\% of the peak-to-peak amplitude as shown in Supplementary Fig. \ref{fig:gaussian_model}. 
According to the law of large numbers, the standard deviation of AWGN is linearly proportional to $\sqrt{n}$, where $n$ is the number of time traces. In this regard, we applied the AWGN to our dataset to simulate the different levels of noise conditions. We also present this noise amplitude in terms of SNR by adopting the equation (\ref{eq:dB}) from the speech signal processing field.

\begin{align}\label{eq:dB}
	\text{SNR} = 10\log_{10}\left(\frac{P(S)-P(N)}{P(N)}\right),
\end{align}
where $P(S)$ and $P(N)$ are the power of signal and noise, respectively.
\begin{figure}
  \centering
  \includegraphics[width=1.0\linewidth]{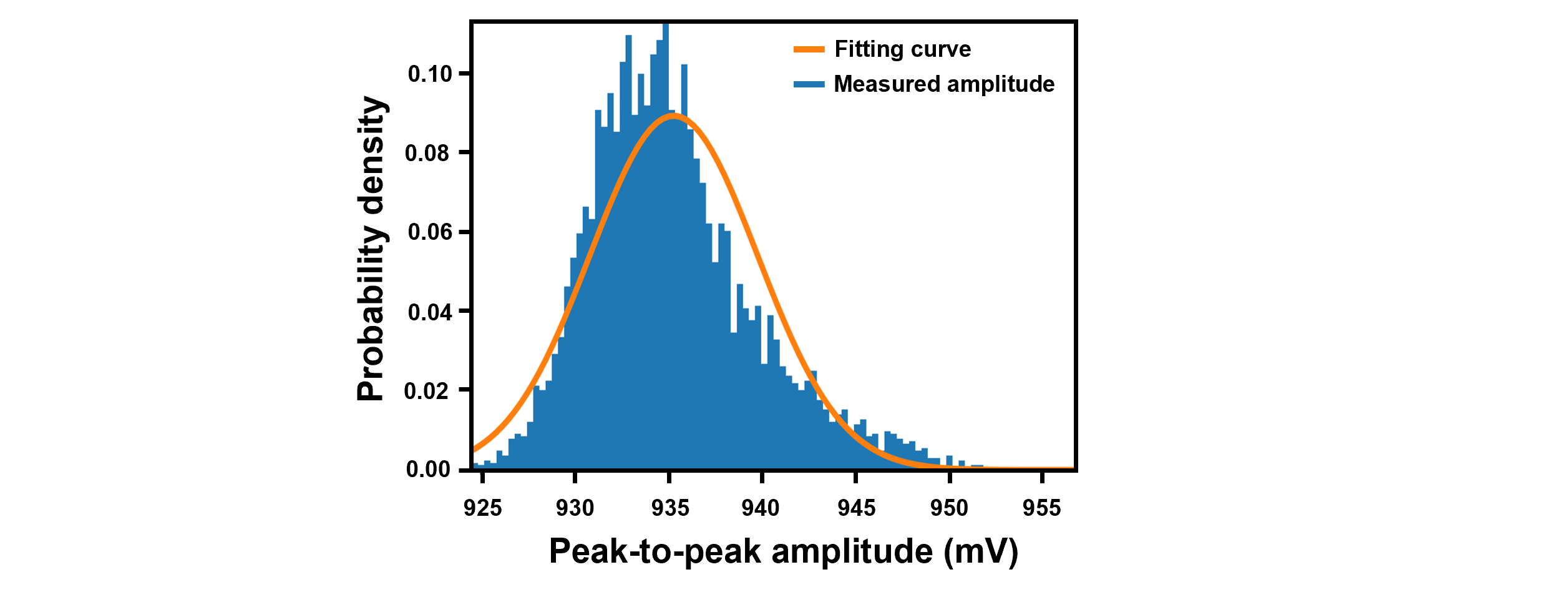}
  \captionsetup{font={small,stretch=1.25}}
  \caption{\textbf{THz-TDS system noise distribution.} The noise distribution in our THz-TDS system follows the Gaussian distribution. The noise standard deviation is 0.48\% of the mean of the peak amplitude. Based on this information, we can infer the noise level in different average number configuration.}
  \label{fig:gaussian_model}
\end{figure}

\subsection*{\large{Supplementary Note 11: Asynchronized Optical Sampling (ASOPS) THz-TDS System}}

The conventional THz time-domain spectroscopy (THz-TDS) system is essential for many THz applications, such as remote sensing, material characterization, imaging, and defect inspection. 
As THz-TDS systems are capable of sampling ultrafast THz signals in the time domain, THz-TDS systems can provide more information, such as time delay, spectral phase, and spectral amplitude \cite{jepsen2011terahertz}. 
A conventional THz-TDS system utilizes a mechanical delay-line stage to introduce a time delay between two split femtosecond laser beams: the so-called pump beam and probe beam \cite{neu2018tutorial}. The pump beam is coupled to the THz photoconductive antenna emitter to generate the THz radiations. The probe beam is fed to the THz photoconductive antenna detector to gate received THz signals \cite{burford2017review,jepsen2011terahertz}. 
With the accumulated time delay between two beams, a THz-TDS system can profile the THz time-domain signal, where the effective sampling frequency is typically tens of THz, which is the reciprocal of the time delay \cite{neu2018tutorial}. 
However, the sampling time and the quality of the time-domain signal are severely limited by the mechanical delay line stage condition, such as maximum moving velocity, position resolution, and operational stability. 
To address the limitation, the asynchronized optical sampling (ASOPS) THz-TDS system is developed to increase the sampling frequency. Additionally, the ASOPS THz-TDS system is more stable due to the elimination of the mechanical delay line stage interference \cite{janke2005asynchronous}. 
By replacing the mechanical delay line stage with the two asynchronized femtosecond lasers, the time delay between the probe beam and the pump beam can be accumulated in a short time. 
More specifically, the sampling time of a single time trace, $t_\text{sample}$, can be decreased down to within millisecond timescale. Additionally, $t_\text{sample}$ is related to the repetition rate of the lasers as
\begin{equation}
\begin{aligned}
t_\text{sample}=\frac{\frac{1}{f^A_\text{rep}}\times\frac{1}{f^A_\text{rep}}}{\frac{1}{f^A_\text{rep}}-\frac{1}{f^A_\text{rep}+\Delta f}}\cong\frac{1}{\Delta f}, \nonumber
\end{aligned}
\end{equation}
where $f^A_\text{rep}$ and $\Delta f$ are the repetition rate of the pump beam laser and repetition rate difference between the two femtosecond lasers. 
In our ASOPS THz-TDS system, the repetition rate of the pump beam laser is 100 MHz. To achieve the high time-domain sampling resolution and the fast sampling time simultaneously, the repetition rate difference is set as 10 Hz, delivering 5 fs sampling resolution and 100 ms sampling time.
However, under this configuration, the size of the full single-time trace is considerably large, requiring high data transfer rate and large storage space. To prevent those issues, we have only taken 100 ps segment of the single time trace and set the sampling rate of the DAQ card as 20 MHz. Accordingly, our ASOPS THz-TDS system provides 0.01 THz resolution in the spectrum and 5 fs time-domain sampling resolution.

\subsection*{\large{Supplementary Note 12: Frequency Band Selection}}

The detectable bands of the spectra are between 0.2 THz and 1.75 THz due to physical and hardware limitations \cite{tani2002generation, preu2011tunable}; the THz spectral signal below 0.2 THz suffers from the physical and instrumental limitations, such as photoconductive material carrier lifetime, bandwidth, and sensing area of the THz photoconductive antennas; the THz spectral signal above 1.75 THz decays to the system noise level due to the absorption of the tablets. Consequently, we have only extracted THz bands between 0.2 THz and 1.75 THz as our valid spectra for the further unmixing experiments.

\subsection*{\large{Supplementary Note 13: Nonnegative Matrix Factorization (NMF)}}
The nonnegative matrix factorization (NMF) is one of the commonly used methods to recover material signatures in the blind sense. 
To address the non-convexity of NMF, the result of NMF is taken from one of the 10 trials under different initialization based on the best root-mean-squared error (RMSE). The RMSE, $D$, is defined as:
\begin{equation}
    \begin{aligned}
    D=\frac{\left\|\bm{X}-\bm{WH} \right\|_F}{\sqrt{nm}}, \nonumber
    \end{aligned}
\end{equation}
where $\mathbf{X} \in\mathbb{R}^{n\times m}$, $\mathbf{W} \in \mathbb{R}^{n \times k}$, and $\mathbf{H} \in \mathbb{R}^{k \times m}$ are the target, signatures matrix, and abundance matrix, respectively. In the experiment, $k$ equals to 3 and 5 in the ternary and quinary cases, respectively. Additionally, we adopt the naive NMF in equation (\ref{eg:NMF}) since validity of the commonly-used regularizers, such as $L^2$ norm or $L^1$ norm, have not yet been verified with reasonable physical meaning. 
The alternating least square method is used to solve this NMF optimization problem:
\begin{equation}
\begin{aligned} \label{eg:NMF}
    \min_{\ensuremath{{\bm W}},~\!\ensuremath{{\bm H}}}~&~\left\| \ensuremath{{\bm X}}-\ensuremath{{\bm W}}\ensuremath{{\bm H}}\right\|_F
	\\
	\textrm{s.t.}~&~
	{ 
	\ensuremath{{\bm W}}\geq\bm 0,~
	\ensuremath{{\bm H}}\geq\bm 0.}
\end{aligned}
\end{equation}

\subsection*{\large{Supplementary Note 14: Linear Mixing Model}}
Since the sum-to-one criterion derived in the ``Methods" section is based on the assumption where interface power loss is relatively small compared to material absorption. 
To validate the assumption in practical cases, we have measured the spectra of three mixture tablets with different mixing approaches as shown in Supplementary Fig. \ref{fig:permutation}. 
The three tablets are composed of the same amount of substances. As shown in Supplementary Fig. \ref{fig:freq_permutation}, the deviation among three measured spectra by different mixing approaches is only less than 2 dB. Based on the experiment, it concludes that the spectrum is not severely affected by the mixing approaches.

In addition to the mixing approaches, the dynamic range of the THz-TDS system affects the accuracy of linear mixing model on the material absorption spectra since the largest detectable absorption coefficients $\alpha_\text{max}(f)$ are determined by the THz-TDS system dynamic range \cite{jepsen2005dynamic}. More specifically, the performance of HYPERION will be decreased when the material absorption coefficients are larger than the $\alpha_\text{max}(f)$. To this sense, it is important to compare the $\alpha_\text{max}(f)$ of our ASOPS THz-TDS system to the pure substance absorption spectra. As shown in Supplementary Fig. \ref{fig:alpha_max}, in the range between 1.3 THz and 1.5 THz, the material absorption coefficients of D-lactose monohydrate and D-glucose excess the $\alpha_\text{max}(f)$. Except this region, the material absorption spectra are less than the $\alpha_\text{max}(f)$. Although the measured absorption coefficients are not accurate in this region, HYPERION can still deliver the acceptable unmixing performance due to use of the material absorption information from all measured frequency bands.

\begin{figure}
    \centering
    \includegraphics[width=1.0\linewidth]{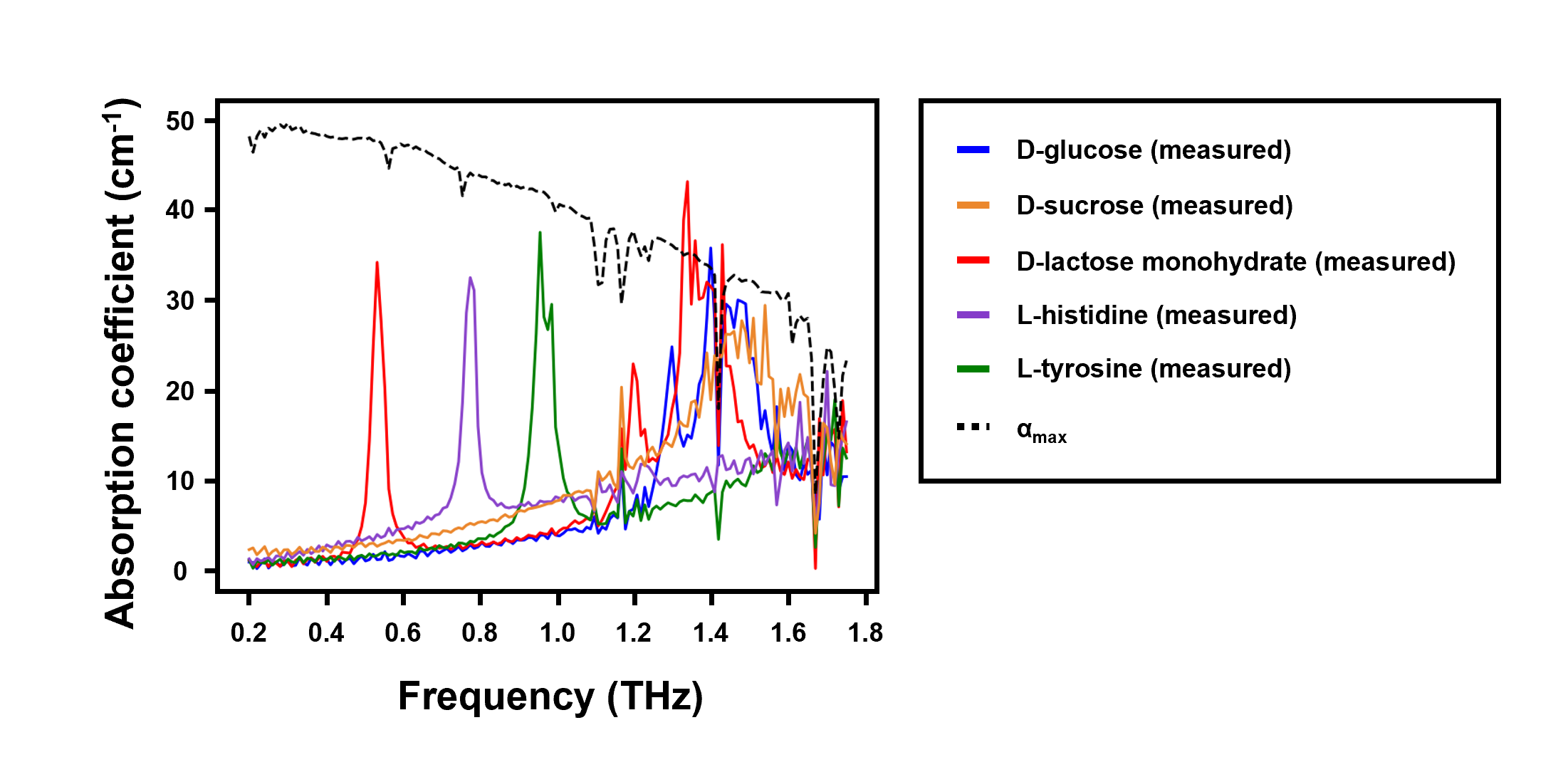}
    \captionsetup{font={small,stretch=1.25}}
    \caption{The comparison among absorption spectra of five materials and the largest detectable absorption spectra $\alpha_{max}$ over the range of 0.2 to 1.75 THz.}
    \label{fig:alpha_max}
\end{figure}

\subsection*{\large{Supplementary Note 15: Composition Estimation Comparison of HYPERION, NMF, HMFA, nICA, and SPA}}

We compared the material composition estimation by the material absorption spectra unmixed from different methods, including HYPERION, NMF, HMFA, nICA, and SPA. The configuration of the test set is same as in the ``Application'' section. As shown in Supplementary Table. \ref{fig:composition_estimation}, HYPERION delivers superior performance of estimating material composition since HYPERION unmixs the accurate material absorption spectra. HMFA and nICA have the much inferior performance on material estimation since the unmixed material absorption spectra are much suppressed in magnitude. The unmixed material spectra by the quinary dataset without pure substances are used to restore the chemical compositions of the 15 tablets in the test set by the convex optimization problem in equation (\ref{eq:application_supp}).

\begin{align} \label{eq:application_supp}
	\min_{\bm{r}_i\in\mathbb{R}^{3}}~&~\left\| \bm{x}_i - \bm{A}\bm{r}_i\right\|_1
	\\
	\textrm{s.t.}~&~
	\ \bm{r}_i \geq 0,
	\ \bm{1^T}\bm{r}_i = 1, 
\end{align}

\begin{table}
  \centering
  \captionsetup{font={small,stretch=1.25}}
  \caption{The composition estimation comparison between HYPERION, HMFA, NMF, SPA and nICA.}
  \includegraphics[width=0.95\linewidth]{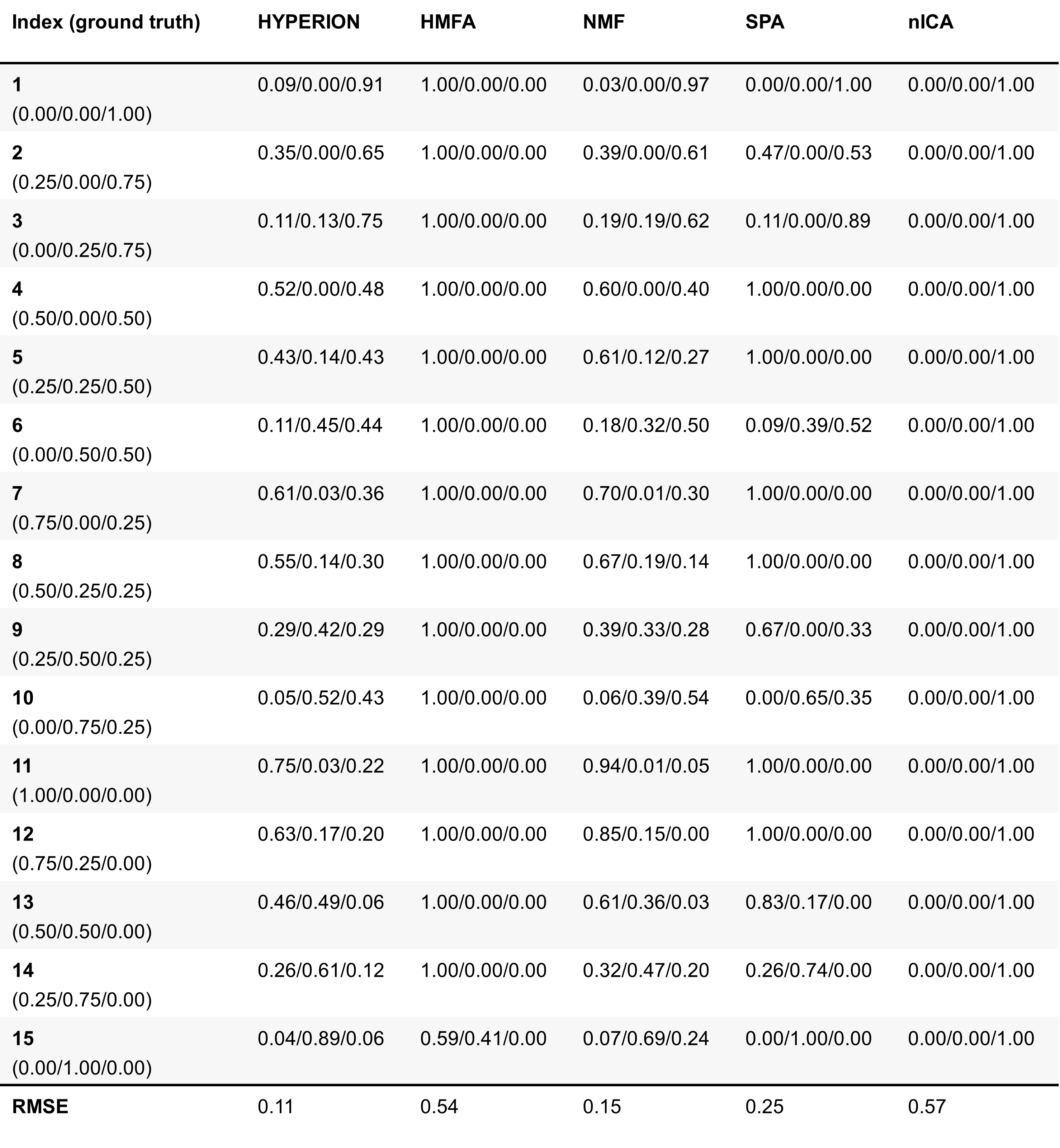}
  \label{fig:composition_estimation}
\end{table}

\subsection*{\large{Supplementary Note 16: Grain Size of Pure Substances and the Scattering Effect}}

Here, we will investigate the effect of the scattering on the unmixed spectra results. Since the effect of scattering is proportional to the grain size of the tablet \cite{bandyopadhyay2007effects}, an optical microscopy with 1.25 numerical aperture (NA) is utilized to measure the grain sizes of different tablets. In the measurement, all chemicals follow the same grounding approach (see ``Method'' for the details) and are placed on the standard microscopy slide. For each chemical, the grain size is determined by taking the average of the grain size of 7 randomly selected grains. As shown in Supplementary Table. \ref{fig:grain_size}, the grain sizes of all chemicals fall in 1 µm range. Thus, the scattering effect in different chemicals falls in the same scale. To this sense, we can model the measured material absorption as the addition of true material absorption and the scattering effect as in equation (\ref{eq:scattering}).

\begin{align} \label{eq:scattering}
    u_{msr}(f) = u_{abs}(f) + u_{sct}(f),
\end{align}
where $u_{msr}$, $u_{abs}$, and $u_{sct}$ are the measured material absorption, material absorption and the scattering effect, respectively. Since the grain sizes of different chemicals fall in the same range, $u_{sct}(f)$ is considerably identical for different chemicals. In HYPERION, this constant shift $u_{sct}(f)$ will be calibrated in the affine fitting step and added to the unmixed spectrum in the transformation between the preconditioned space and original space. Thus, the scattering effect will not affect the unmixing performance of HYPERION when the grain sizes of different chemical fall in the same scale.

\begin{table}
  \centering
  \captionsetup{font={small,stretch=1.25}}
  \caption{The grain size of five pure substances. The measurement is performed by a optical microscopy with numerical aperture (NA) of 1.25.}
  \includegraphics[width=0.95\linewidth]{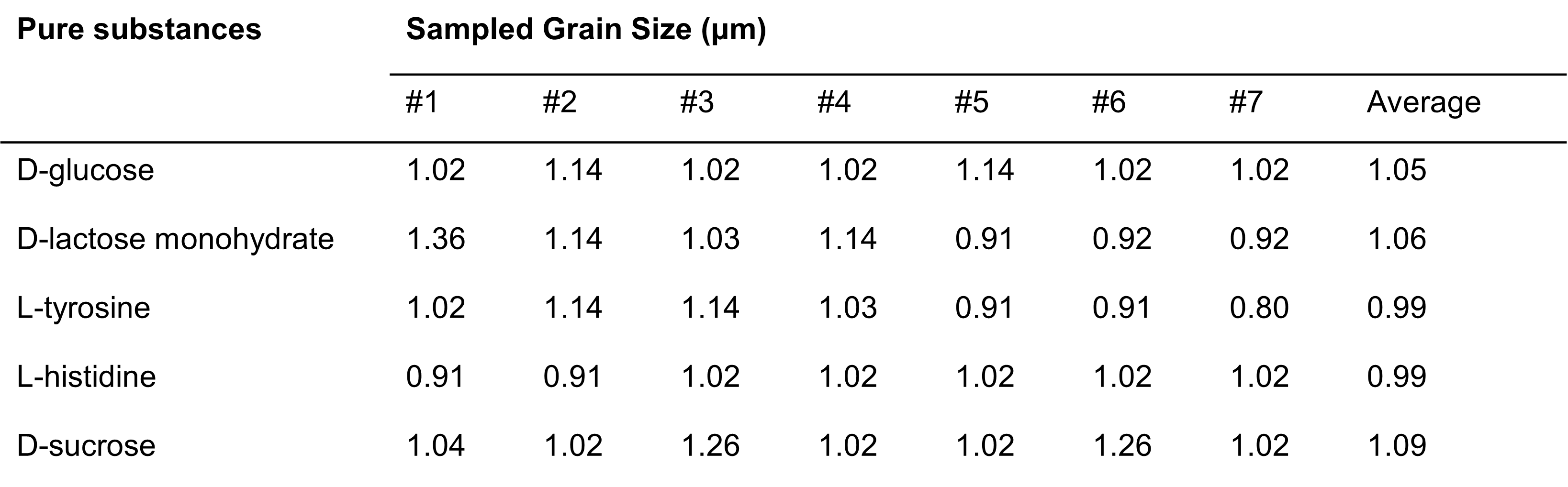}
  \label{fig:grain_size}
\end{table}

\FloatBarrier
\begin{figure}
  \centering
  \includegraphics[width=1.0\linewidth]{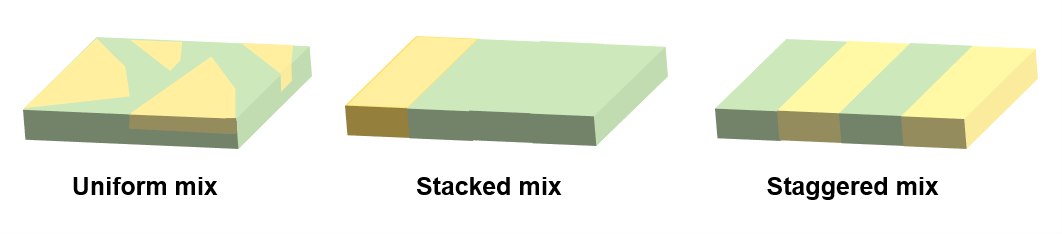} 
  \captionsetup{font={small,stretch=1.25}}
  \caption{\textbf{Three different permutation approaches.} The 1$^\text{st}$ way is to mix chemicals randomly, the 2$^\text{nd}$ way is to mix chemicals in order, and the 3$^\text{rd}$ way is to mix chemicals in a staggered manner.}
  \label{fig:permutation}
\end{figure}

\begin{figure}
  \centering
  \includegraphics[width=1.0\linewidth]{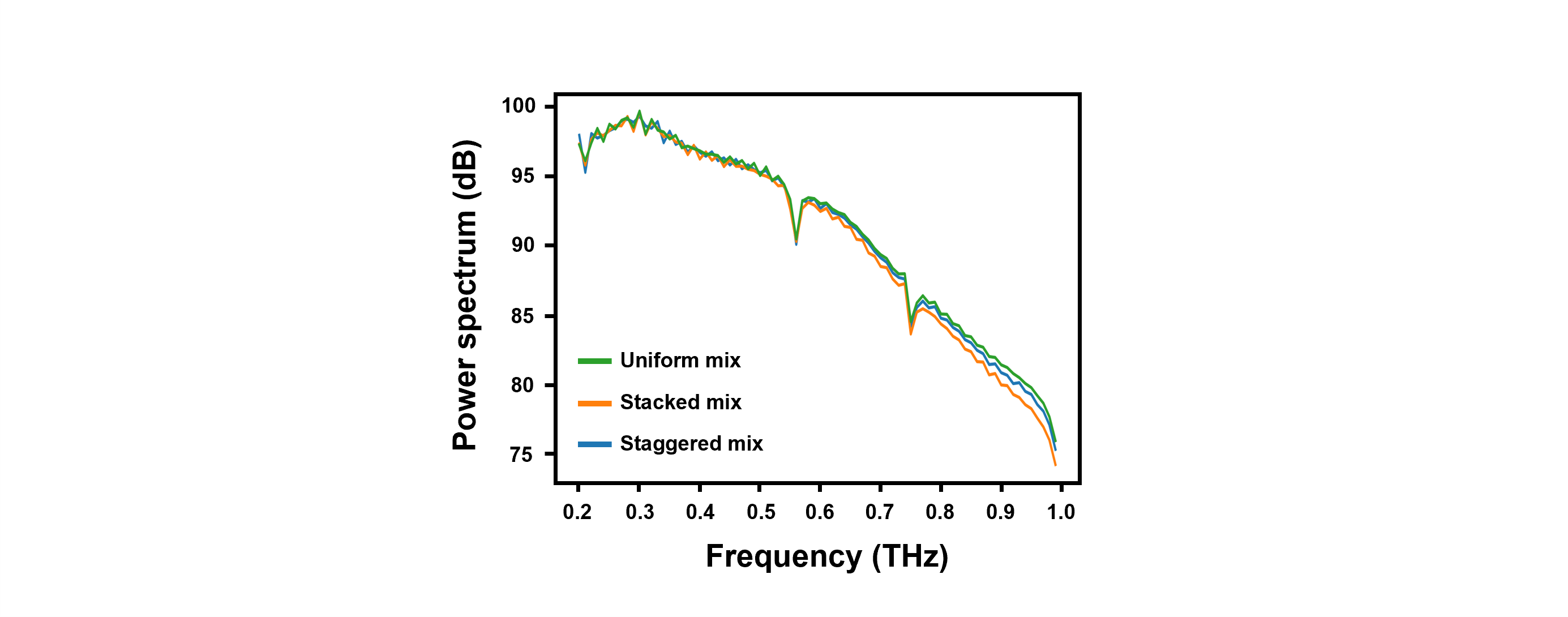}
  \captionsetup{font={small,stretch=1.25}}
  \caption{\textbf{The THz frequency-domain spectrum of different permutation approaches.} The amplitude deviation among three different permutation approaches (uniform mix, stack mix, and staggered mix) is less than 2 dB within 0.2 - 1.0 THz.}
  \label{fig:freq_permutation}
\end{figure}

\FloatBarrier

\begin{table}
  \centering
  \captionsetup{font={small,stretch=1.25}}
  \caption{The quantitative comparison between the HYPERION and NMF on five substances (HYPERION/NMF) under different noise levels. Performances of HYPERION and NMF are evaluated by the root mean square error (RMSE). Superior values are labeled in bold and are highlighted with different colors (HYPERION: orange; NMF: blue).}
  \includegraphics[width=0.95\linewidth]{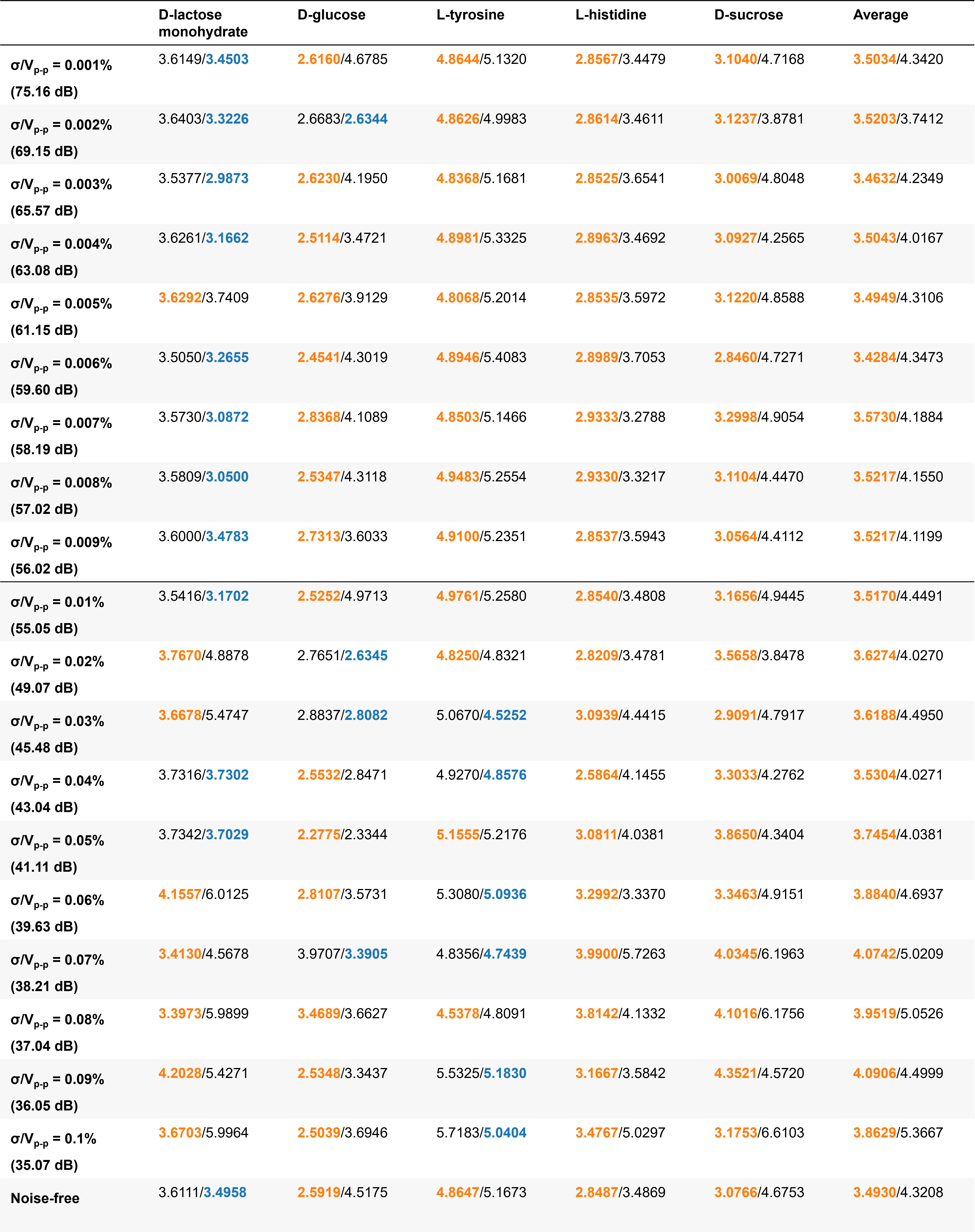}
  \label{fig:table_hu_vs_nmf}
\end{table}

\begin{table}
  \centering
  \captionsetup{font={small,stretch=1.25}}
  \caption{The quantitative comparison between the HYPERION and HMFA on five substances (HYPERION/HMFA) under different noise levels. Performances of HYPERION and HMFA are evaluated by the root mean square error (RMSE). Superior values are labeled in bold and are highlighted with different colors (HYPERION: orange; HMFA: blue).}
  \includegraphics[width=0.95\linewidth]{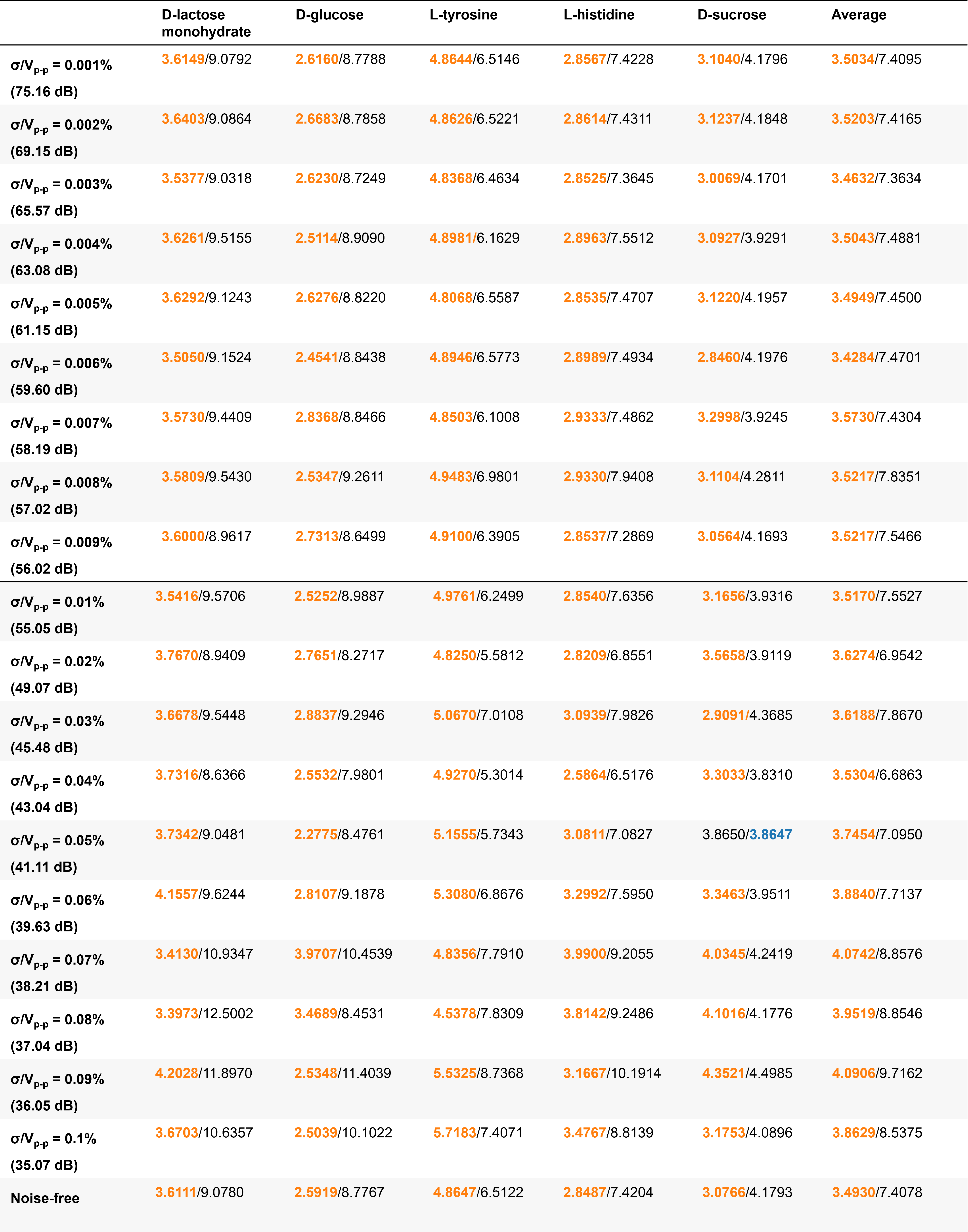}
  \label{fig:table_hu_vs_hmfa}
\end{table}

\begin{table}
  \centering
  \captionsetup{font={small,stretch=1.25}}
  \caption{The quantitative comparison between the HYPERION and nICA on five substances (HYPERION/nICA) under different noise levels. Performances of HYPERION and nICA are evaluated by the root mean square error (RMSE). Superior values are labeled in bold and are highlighted with different colors (HYPERION: orange; nICA: blue).}
  \includegraphics[width=1.0\linewidth]{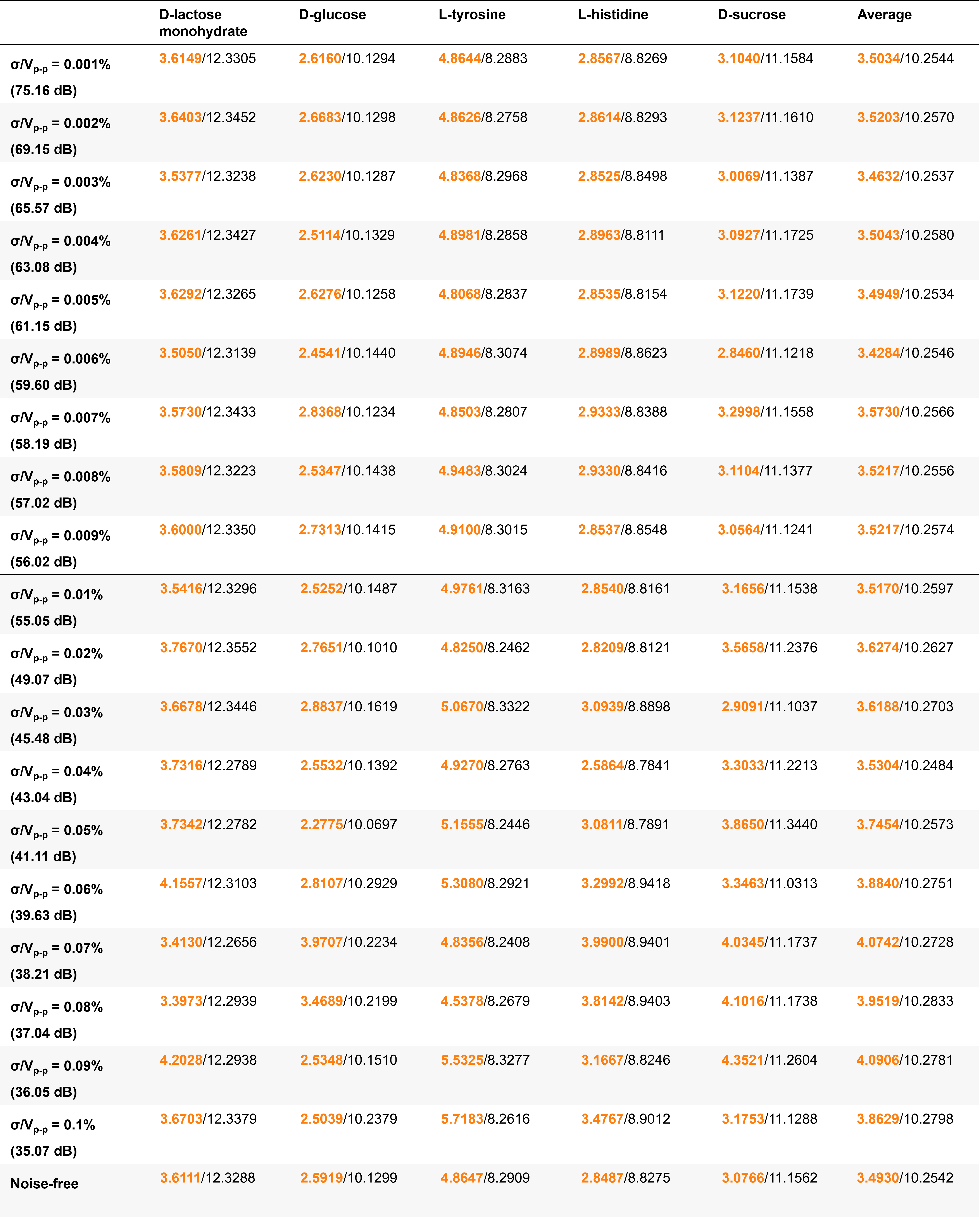}
  \label{fig:table_hu_vs_nica}
\end{table}

\begin{table}
  \centering
  \captionsetup{font={small,stretch=1.25}}
  \caption{The quantitative comparison between the HYPERION and SPA on five substances (HYPERION/SPA) under different noise levels. Performances of HYPERION and SPA are evaluated by the root mean square error (RMSE). Superior values are labeled in bold and are highlighted with different colors (HYPERION: orange; SPA: blue).}
  \includegraphics[width=1.0\linewidth]{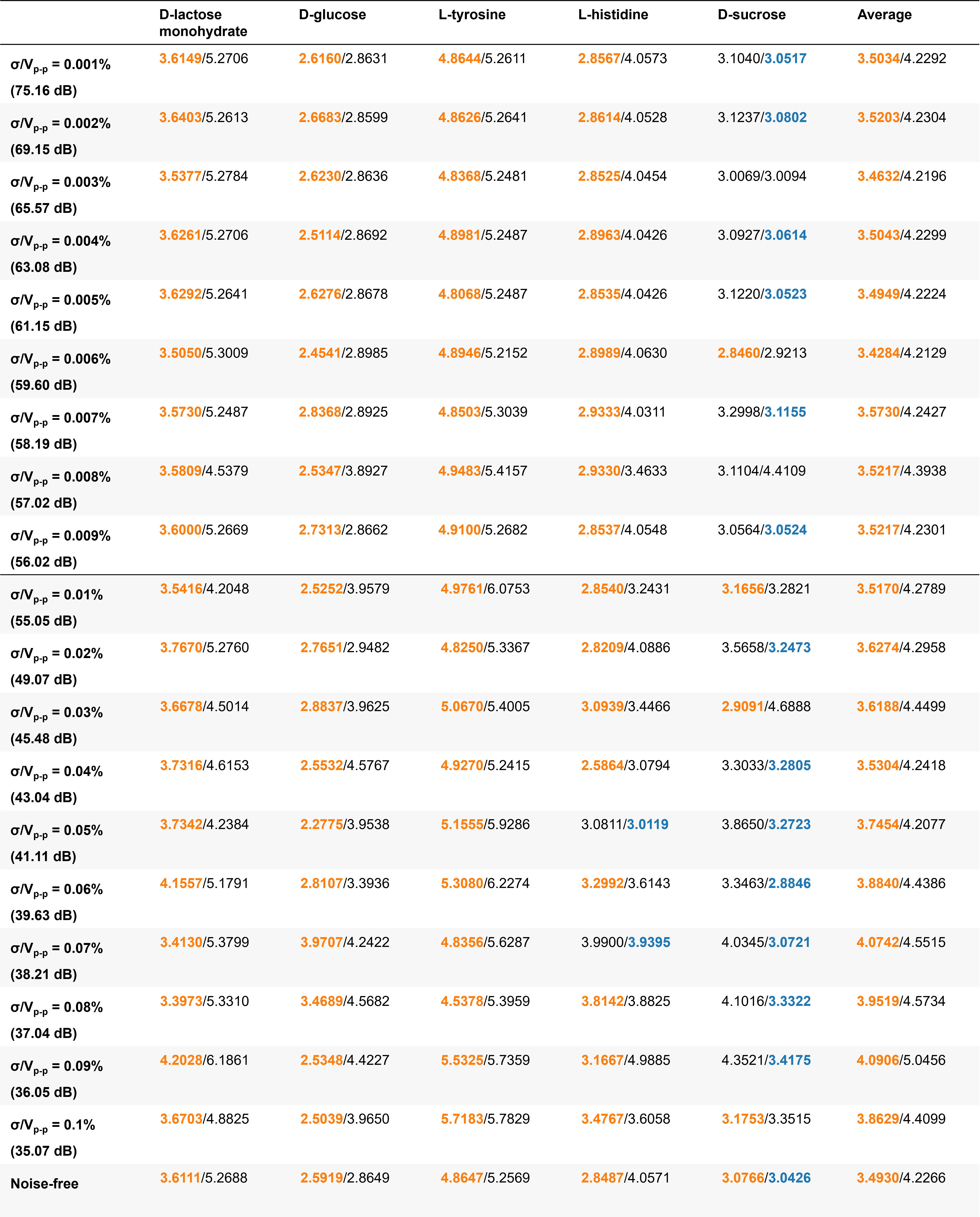}
  \label{fig:table_hu_vs_spa}
\end{table}

\begin{table}
  \centering
  \captionsetup{font={small,stretch=1.25}}
  \caption{Tablet thickness in the quinary dataset.}
  \includegraphics[width=1.0\linewidth]{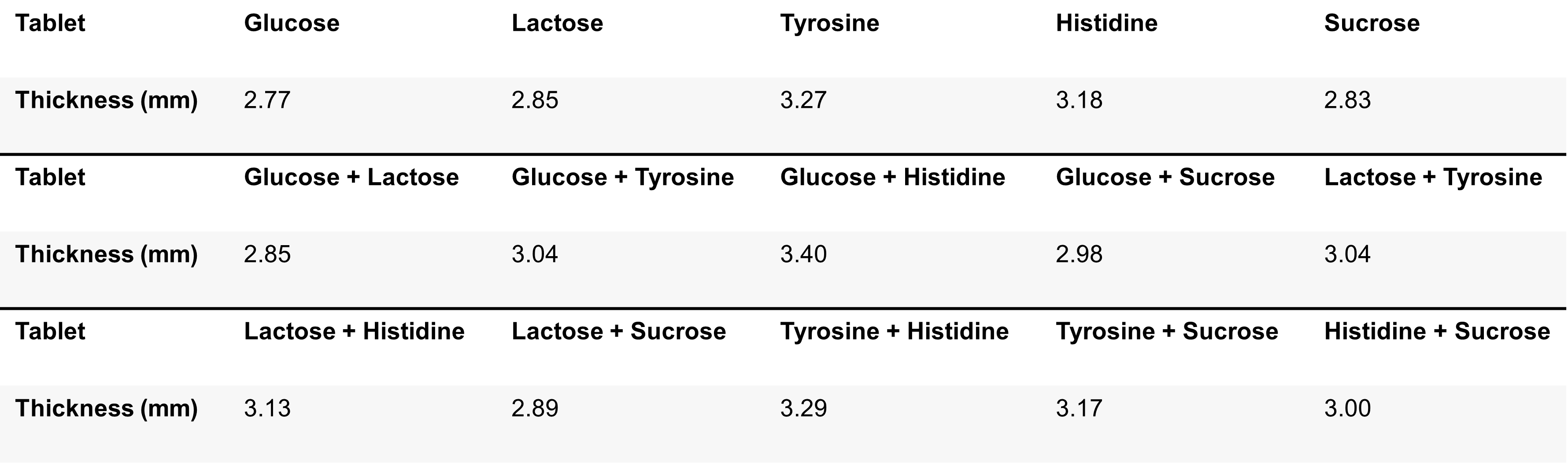}
  \label{fig:tablet_thickness}
\end{table}

\FloatBarrier

\begin{table}
  \centering
  \captionsetup{font={small,stretch=1.25}}
  \caption{The convergence time of HYPERION, HMFA, NMF, SPA, and nICA.}
  \includegraphics[width=1.0\linewidth]{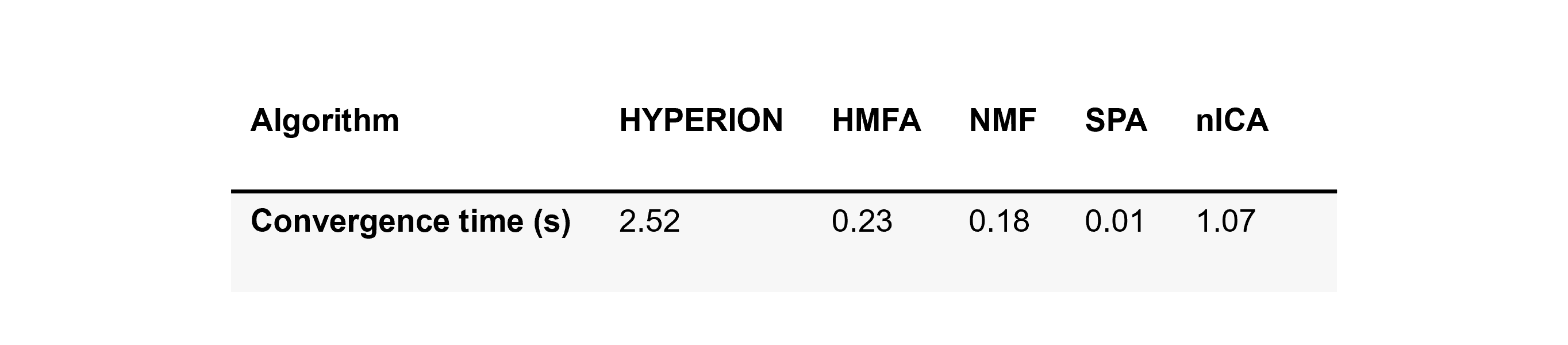}
  \label{fig:convergence_time}
\end{table}

\FloatBarrier

\printbibliography

\end{document}